\documentclass[journal]{IEEEtranTIE}
\usepackage{graphicx}
\usepackage{cite}
\usepackage{picinpar}
\usepackage{amsmath,amsfonts,amssymb,amscd}
\usepackage{amsthm}
\usepackage{array}
\usepackage{url}
\usepackage{flushend}
\usepackage{colortbl}
\usepackage{nomencl}
\usepackage{soul}
\usepackage{multirow}
\usepackage{pifont}
\usepackage{color}
\usepackage{mathtools}
\usepackage{blkarray}
\usepackage{alltt}
\usepackage[hidelinks]{hyperref}
\usepackage{enumerate}
\usepackage{siunitx}
\usepackage{breakurl}
\usepackage{epstopdf}
\usepackage{arydshln}
\usepackage[shortlabels]{enumitem}
\usepackage{pbox}
\usepackage{makecell}
\usepackage[labelformat=simple]{subcaption}
\usepackage{tabularx}
\usepackage{booktabs}
\usepackage{hhline}
\usepackage{dblfloatfix}
\usepackage{float}
\usepackage[utf8]{inputenc}
\usepackage{longtable} 
\usepackage{enumitem}  
\usepackage{acronym}

\newcolumntype{?}{!{\vrule width 1pt}}

\let\oldhref\href
\renewcommand{\href}[2]{\oldhref{#1}{\hbox{#2}}}
\begin{document}
\title{ Stability Analysis in IBR-Dominated Grids: Model-Based and Data-Driven Methodologies }
%

 \author{
 	\vskip 1em
 	{\color{black}
 	Federico~Cecati, \emph{Member, IEEE},
 	Yang~Wu, \emph{Member, IEEE},
     Florian~D\"orfler \emph{Fellow, IEEE},
     Marco~Liserre, \emph{Fellow, IEEE}, 
 	Xiongfei~Wang, \emph{Fellow, IEEE},
    Elisabetta~Tedeschi, \emph{Senior Member, IEEE},	
 	} 
}
%
%
%
%
%
%
%

%
%
%
%
%
%
%
\maketitle
\begin{abstract}
    Stability issues in converter-dominated grids have a multifold nature, where static, small-signal and large-signal phenomena can influence each other. Capturing the interplay among these different stability aspects requires a rethinking of conventional modeling and analysis tools for power systems. As a further complication, Intellectual Property protection among manufacturers and system operators makes detailed converters and grid models unattainable. This motivates the use of data-driven approaches for stability analysis and control, nevertheless achieving an insightful representation of static, small-signal and large-signal phenomena from data still leaves many open questions.
    This paper discusses model-based and data-driven methodologies for stability analysis of converter-dominated grids. The instability mechanisms of static, small-signal and large-signal phenomena are conceptually explained, and the capability of different modeling and analysis tools to investigate them is examined. The effectiveness of data-driven methodologies to analyze instability phenomena in their multifold nature is debated, and the challenges on data completeness, model generalization under untrained conditions, physical interpretability, and uncertainty awareness are discussed.
\end{abstract}
\begin{IEEEkeywords}
Large-signal stability, Small-signal stability, Data-driven, Model-based, Inverter-Based Resources
\end{IEEEkeywords}
\markboth{IEEE TRANSACTIONS ON POWER ELECTRONICS}%
{}
\definecolor{limegreen}{rgb}{0.2, 0.8, 0.2}
\definecolor{forestgreen}{rgb}{0.13, 0.55, 0.13}
\definecolor{greenhtml}{rgb}{0.0, 0.5, 0.0}
%
%
%
%
%
%
%
%
\maketitle
\section*{List of Acronyms}
\begin{acronym}[IEEEtran] 
    \acro{CLM}{Current Limitation Mode}
    \acro{EMT}{Electromagnetic Transient}
    \acro{FRT}{Fault-Ride Through}
    \acro{IBR}{Inverter-Based Resource}
    \acro{IP}{Intellectual Property}
    \acro{LOS}{Loss Of Synchronism}
    \acro{MIMO}{Multi-Input Multi-Output}
    \acro{NOM}{Normal Operating Mode}
    \acro{OP}{Operating Point}
    \acro{PCC}{Point of Common Coupling}
    \acro{PI}{Proportional Integral}
    \acro{PINN}{Physics-Informed Neural Network}
    \acro{PRBS}{Pseudo-Random Binary Sequence}
    \acro{SCR}{Short Circuit Ratio}
    \acro{SG}{Synchronous Generator}
    \acro{VF}{Vector Fitting}
\end{acronym}


\section{Introduction}
%
%
In the last decades, the role of power electronics in electric grids has been elevated from a sole power conversion stage for small renewable generation units to a smart and multifunctional interface for multi-MW power plants with grid stabilizing capabilities \cite{paolone2020fundamentals,blaabjerg_intro}. With the increasing adoption of grid forming control, Inverter-Based Resources (IBRs) are taking more and more responsibility in establishing the voltage and regulating the grid frequency, taking over the traditional role of Synchronous Generators (SG) \cite{roberto_overview_gridforming,ngeso_gfm_2025,UNIFI_GFM_Specs_V3}. \par 
\begin{table}[tb]
	\centering
	\caption{Overview of power system instability phenomena.}
	\label{tab:static_dyn_linear_analysis}
	\begin{tabular}{ ? l ? l ? }
		\Xhline{2\arrayrulewidth}
		\textbf{Type of instability} & \textbf{Grid phenomena it involves}  \\ 
        \Xhline{2\arrayrulewidth}
        Large-signal & - Exceeding maximum power transfer capability \\
        static instability  & - Reaching static voltage stability limit  \\
         \centering \textit{or}  & - Post-fault equilibrium point non-existence \\
        Static instability   & - Lack of long-term generation/demand balance \\
        \  & - Inconsistent power set-points \\
        \  & - Voltage collapse \\
        \  & - Long-term bus voltage limit violation \\
    	\Xhline{2\arrayrulewidth}
        Small-signal & - Sub-synchronous oscillations \\
        instability &  - Inter-area oscillations \\
        \  & - Converter-driven oscillatory instability \\
        \  & - Harmonic instability \\
        \  & - Resonance instability \\
        \Xhline{2\arrayrulewidth}
        Large-signal & - Dynamic voltage instability \\
        dynamic instability & - Protection triggering during contingencies \\
          & - DC-link overvoltage during faults \\
          & - Overcoming IBR physical limits during transients \\
          & - Windup phenomena \\
          & - Loss of Synchronism during phase jumps \\
        \Xhline{2\arrayrulewidth}
	\end{tabular}
\end{table}
This paradigm shift toward a power electronics-dominated grid led to the emergence of new and diverse stability issues \cite{taskforce_stab_revised}, increasingly observed in recent blackouts, including undamped oscillations and voltage regulation problems (e.g. Spain 2025 \cite{spanish_blackout_finalreport}), low inertia (e.g. South Australia 2016 \cite{australia_blackout,paolone2020fundamentals}), and large-scale tripping of renewable power plants due to loss of synchronism (e.g. Texas Odessa \cite{TexasOdessa}). The mechanisms behind converter-driven grids instability events are highly complex and \emph{multifold} \cite{taskforce_stab_revised}, i.e. they can simultaneously involve small-signal phenomena (e.g. undamped oscillations), large-signal phenomena (e.g. loss of synchronism due to current limitation) and static phenomena (e.g. insufficient long-term reactive power support) \cite{taskforce_stab_revised,costas_book,florian_lowinertia}. To ensure grid stability under all points of view, recent grid codes -- especially with the emerging grid forming converters -- are updating IBR functional requirements considering concurrently multiple aspects, e.g. oscillation damping capability, voltage regulation policies, grid supporting functionalities, Fault-Ride Through (FRT) behaviour \cite{ngeso_gfm_2025,entsoe_gfm_phase2_2025,UNIFI_GFM_Specs_V3}. \par 
Analyzing all these diverse aspects and meeting all IBR requirements \emph{simultaneously} is a tough problem, which requires analytic tools able to perform static, small-signal, and large-signal dynamic analyses \cite{fmilano_opensource,cecati_modeling_journal}. A formidable challenge for model-based analysis tools is to capture the \emph{interplay} among different aspects of grid stability. As an example, long-term reactive power regulation in IBRs is commonly used for improving static grid operation, yet it concurrently affects also IBR oscillation damping capability and grid small-signal stability, by changing the grid equilibrium point \cite{cecati_modeling_journal}. Furthermore, large-signal instability, usually considered as a dynamic problem, is often driven by the non-existence of a post-fault equilibrium point, whose assessment is rather a static problem \cite{cecati_phdthesis, xiongfei_fastslow}. \par 
Analyzing IBR-based grid stability, embracing the interrelation between static and dynamic aspects, is further complicated by Intellectual Property (IP) protection among different manufacturers, system operators and other stakeholders \cite{xiongfei_edge}. That makes detailed analytical models of power electronics-dominated grids unattainable, and heavily limits the applicability of conventional model-based methodologies \cite{cecati_interop}. To overcome this issue, data-driven methodologies for stability analysis and control of IBR-dominated power systems are gaining increasing attention \cite{xiongfei_edge,yang_imped_datadriven}. However, capturing the interplay among different aspects of the stability with data-driven methods is even more difficult than with model-based methods, and leaves still nowadays many open questions. \par 
This paper discusses the theoretical foundation, latest developments, and research gaps in model-based and data-driven stability analysis tools for IBR-dominated power systems. \par
\begin{figure}[t!]
	\centering
	\includegraphics[width= \linewidth]{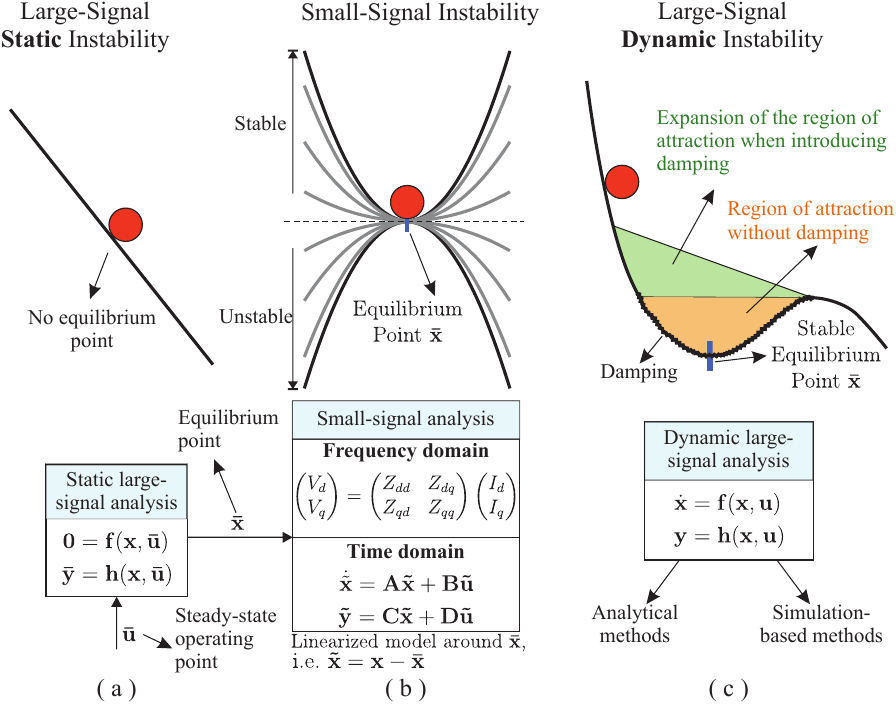}  
	\caption{ Conceptual representation of different stability problems and their analysis tools. (a) Existence of the equilibrium point. (b) Local stability and dynamic properties of the equilibrium point (if existing). (c)   
    Convergence to the equilibrium point with limited region of attraction.  }
	\label{fig:stability-ball}
\end{figure}

 
%
The instability mechanisms of static, small-signal and large-signal phenomena are firstly elaborated (see Table \ref{tab:static_dyn_linear_analysis} and Fig. \ref{fig:stability-ball}). Then, the responsible factors behind these three types of instabilities in IBR-dominated grids are unveiled, focusing on the impact of IBR control system dynamics, IBR physical constraints (e.g. current limitation) and interaction with AC grids and with other IBRs (see Table \ref{tab:physic_constraints}). What emerges from the discussion is that several dynamics and constraints of IBRs and power grids affect different types of stability \emph{simultaneously}. Additionally, for some control dynamics e.g. synchronization, the guidelines for improving different aspects of the IBR stability are even conflicting: slow synchronization improves the small-signal stability \cite{cecati_modeling_journal,xiongfei_fastslow}, nevertheless it deteriorates the large-signal stability \cite{hengwu_trans_PLL, hengwu_trans_pow}. These considerations motivate a profound rethinking of IBR stability analysis tools to offer a more holistic perspective on stability, i.e. be able to study the interplay among static, small-signal and large-signal aspects. \par 
With model-based methods, properly representing the IBR nonlinearity and performing combined static/dynamic stability analyses are of utmost importance to capture this interplay \cite{cecati_modeling_journal}.

%
%
%
%
%
With data-driven methods, in addition to the two aforementioned factors, assessing the completeness and informativeness of the collected data plays a critical role \cite{yang_imped_datadriven}. For small-signal analyses at a single operating point, for example, IBR impedance measurements over a wide frequency range in $d$ and $q$ channels are sufficient \cite{honggong_estim}. However, to capture the interplay between small-signal stability and static equilibrium point, a much larger amount of data across multiple operating points are required \cite{yang_imped_datadriven}. Studying transient stability issues from data is even more difficult, since quantifying the completeness and informativeness of the collected data in presence of complex nonlinearities is challenging \cite{david_hill_SVM,panos_probabilistic_trasstab}. Once data are collected, a further issue is to find a proper model structure and computational method to fit data into models with high accuracy and generalizability \cite{xiongfei_edge,yang_imped_datadriven}. The generalizability of data-driven models for unseen operating conditions or unseen IBR configurations is indeed always limited; transfer learning has been proposed to partially mitigate this issue by embedding domain knowledge into the learning process \cite{xiongfei_edge, Zilio_TransferLearning_2024}. Moreover, purely black-box data-driven models may still violate fundamental physical constraints, and they lack the interpretability of first-principle models familiar to engineers, reducing their trustworthiness in safety-critical applications. \par 
%
%
%
The contributions of this work are summarized as follows:
\begin{itemize}
    \item The static, small-signal and large-signal dynamic stability of IBR-based grids are reviewed in a tutorial manner (see Table \ref{tab:static_dyn_linear_analysis} and \ref{tab:physic_constraints}), and their interdependency is explained and demonstrated through analyses and simulations. 
    \item Nonlinear model-based tools for combined static/dynamic stability analyses are unveiled (see code repository \cite{repository_cecati_nonlinmod}), and their capability to capture the interplay among different stability aspects is examined (see Figs. \ref{fig:interplay1} and \ref{fig:interplay2}).
    \item Existing data-driven approaches for IBR stability analysis are introduced, and their challenges related to data requirements, generalization under untrained conditions, physical interpretability, and uncertainty awareness are discussed (see Table \ref{tab:datadriven_table}).    
    \item Data-driven tools to capture the interplay between static (i.e. IBR operating point) and dynamic stability aspects (e.g. small-signal stability) are presented, and data-driven control applications to IBR-dominated grids are discussed.
\end{itemize}
The paper is organized as follows: Section \ref{sec:model-based} focuses on model-based methods, and Section \ref{sec:data-driven} on data-driven methods and Section \ref{sec:concl} presents the conclusion and future trends.
%

%
%
%
%
%
%
%
%
%
\begin{table*}[tbp]
\centering
\caption{Overview and description of IBR dynamics and physical constraints in relevance to power grid stability.}
\renewcommand{\arraystretch}{1.5}
\footnotesize
\renewcommand{\tabularxcolumn}[1]{m{#1}}
\begin{tabularx}{\textwidth}{
    >{\arraybackslash}m{2.cm} !{\vrule width 1pt}  %
    >{\centering\arraybackslash}m{1.1cm} !{\vrule width 1pt} %
    >{}X        
}
\textbf{Factors that affect system stability} & \textbf{Type of stability} & \textbf{Impacts of these different factors and limitations of traditional model-based stability analysis methods} \\ 
\Xhline{1pt}
\multirow{3}{\linewidth}{\textbf{IBR control system dynamics}} 
& Static stability &    
    \textbullet~ Active power reduction during faults can avoid static instability (e.g. reducing $P^*$ in Fig. \ref{fig:Pdelta_curve})  \cite{gabsuseo_trans_fictious,currentlim_overview}.  \newline 
    \textbullet~ Voltage/reactive power regulation by IBRs, imposed by grid codes, positively impact grid static stability \cite{IEEE_1547-2018}. 
     \\ \cline{2-3}
& Small-signal stability & 
    \textbullet~  Fast synchronization can jeopardize small-signal stability in GFL and GFM (see Fig. \ref{fig:eig_mp}) \cite{PLL_and_currloop,cecati_design_powsynch}. \newline
    \textbullet~  Fast DC-link voltage regulation can jeopardize small-signal stability in GFL and GFM \cite{cecati_modelling_ECCE}. \newline
    \textbullet~  Reactive power control tuning influences the small-signal stability differently in GFM and GFL \cite{harnefors_robdesign,cecati_modeling_journal,cecati_design_powsynch}. \newline
    \textbullet~  Fast inner loops can trigger resonances in harmonic frequency range \cite{harm_stab_concept,LCL_damping_marco}. 
    \\ \cline{2-3}
& Large-signal dynamic stability & 
    \textbullet~  Slow synchronization can jeopardize transient stability during faults and phase jumps (see Fig. \ref{fig:mp_LOS_n}) \cite{graz_regofattr,hengwu_trans_PLL,xiongfei_fastslow}. \newline
    \textbullet~  Slow DC-link regulation jeopardizes transient stability during faults and phase jumps \cite{transstab_dclink,xiongfei_dclink_transstab}. \newline
    \textbullet~  Insufficient or delayed reactive power support can jeopardize dynamic large-signal stability \cite{jingyang_react_transstab}.  \newline 
    \textbullet~  Fast voltage support is necessary to avoid dynamic voltage instability, and is indeed imposed by modern grid codes \cite{florian_reissner_fastGFM} (e.g. NESO and ENTSO-e \cite{ngeso_gfm_2025,entsoe_gfm_phase2_2025}). \\ 
    \Xhline{1pt}
\multirow{3}{\linewidth}{\textbf{Physical constraints of IBR and energy source/storage}} 
& Static stability & 
    \textbullet~  Current limitation affects the existence of post-fault equilibrium points (see blue curves in Fig. \ref{fig:Pdelta_curve}) \cite{linbin_currlim,currlim_expression_VgIM, xavier_transint_rokrok}. \newline
    \textbullet~ Inconsistent power setpoints between energy source and grid-connected converter can cause static instability \cite{juancamilo_transient}.
 \\ \cline{2-3}
& Small-signal stability & 
    \textbullet~  Some current limitation strategies are prone to cause voltage oscillations during faults \cite{gabsuseo_currlim_smallsign}. \newline
    \textbullet~  Energy source/storage constraints limit the capability of IBRs to damp frequency oscillations \cite{verena_heterogeneous}.
 \\ \cline{2-3}
& Large-signal dynamic stability & 
    \textbullet~  Current limitation mode, often triggered during large transients, can cause windup and instability (see Fig. \ref{fig:simul_SCR}) \cite{gabsuseo_overview_currlim}. \newline
    \textbullet~  Current limitation can impair the IBR DC-link regulation capability, causing DC overvoltage and instability \cite{juancamilo_transient}. \newline
    \textbullet~  Overmodulation can occur during inertia provision, due to DC-link undervoltage, potentially causing instability \cite{cecati_overmodulation}.  \newline
    \textbullet~  IBR overcurrent can be extended through semiconductor thermal monitoring, improving transient stability \cite{cecati_transient_MMC,fabio_mandr_temperature}.  \newline 
    \textbullet~  Limited response speed and energy/power availability of the energy source/storage can jeopardize stability \cite{PMSM_gridforming, electrolyzer_grfrm, flywheel_grfrm}.
    \\ 
    \Xhline{1pt}
\multirow{3}{\linewidth}{\textbf{Interactions with the AC grid dynamics and with the other IBRs' dynamics}} 
& Static stability & 
\textbullet~  SCR and $R/X$ ratio influence maximum power transfer capability, IBR equilibrium point, and static instability \cite{hengwu_trans_PLL}. \newline 
\textbullet~  IBR FRT strategies and power network topology influence static stability during large faults \cite{fmilano_transient,cecati_phdthesis}.  
\\ \cline{2-3}
& Small-signal stability & 
\textbullet~  Weak grid condition can jeopardize GFL-IBRs stability, while stiff grid compromises GFM-IBRs stability \cite{jason_PLL,roberto_synchro_parall}. \newline
\textbullet~  Interconnection to a weak grid is more stable in the case of GFM-IBRs than GFL-IBRs \cite{xiongfei_fastslow}. \newline 
\textbullet~  SCR and $R/X$ ratio metrics provide limited insight on small-signal phenomena, requesting for new and more informative metrics \cite{strenght_gSCR,linbin_strenght}. \newline 
\textbullet~  Inertial response of the grid affects frequency oscillations, yet it is neglected in Thevenin equivalent grid models \cite{markovic_lowinertia}.  \newline 
\textbullet~  Interactions among different IBRs in the power network can create low-damped oscillations \cite{cecati_interop,yicheng_interoperab}.  \newline 
\textbullet~  Improper IBR control tuning may cause resonances that can be easily excited by the rest of the power system \cite{cecati_modeling_journal}.
 \\ \cline{2-3}
& Large-signal dynamic stability & 
\textbullet~  AC grid strength highly affects the onset of Loss of Synchronism (LOS) phenomena (see Fig. \ref{fig:simul_SCR}) \cite{hengwu_trans_PLL,transstab_nonlindamping}. \newline
\textbullet~ AC grids dynamics influences transient stability in a wide frequency range, nevertheless most of analytical proofs model the grid as a pure static reactance \cite{hengwu_trans_PLL,transstab_nonlindamping}. \newline 
\textbullet~  Interaction among diverse IBRs can cause transient instability, yet scalable methods for its analysis are missing \cite{fmilano_transient}.
\\ \cline{2-3}
\Xhline{1pt}
\end{tabularx}
\label{tab:physic_constraints}
\end{table*}
%
%
%
%
%
%
%
%
%
%
%
%
%
%
%
%
%
%
%
%
%
\section{Model-based methodologies for stability analysis of IBR-dominated grids} \label{sec:model-based}

Small-signal and large-signal instability phenomena typical of IBR-based grids have substantially different natures, as illustrated in Fig. \ref{fig:stability-ball} with the analogy of a rolling ball on a curved surface, and are studied through different mathematical tools \cite{cecati_phdthesis,kundur_book}. A nonlinear state-space model of the IBR in a synchronous reference frame is considered (see bottom of Fig. \ref{fig:stability-ball}), where $\mathbf{x}$ is the state, $\mathbf{u}$ the input that contains both disturbances (e.g. grid voltage) and control set-points (e.g. active/reactive power references), and $\mathbf{y}$ is the output. 

%
%
%
%
Large-signal static instability (or simply \emph{static instability}) is the root cause of several grid disruptive events -- described in Table \ref{tab:static_dyn_linear_analysis} -- and is characterized by the absence of a steady-state equilibrium point for the IBR, as in Fig. \ref{fig:stability-ball}(a). Given a constant $\mathbf{u}=\mathbf{\bar{u}}$ which defines the IBR operating point, the non-existence of the equilibrium point $\mathbf{\bar{x}}$ is linked with the unsolvability of IBR state-space equation in static condition ($\dot{\mathbf{x}}=\mathbf{0}$), shown at the bottom of Fig. \ref{fig:stability-ball}(a) and denominated \emph{static equation}.

%
%
%
%
%
%
%
%
%
%
%
%
%
%
%
%
%
%
%
%
%
%
%
%
%
%
%
%
%
%
%
%
%
%
In the case of existence of an equilibrium point, small-signal stability analysis (see Fig. \ref{fig:stability-ball}(b)) assesses the IBR local stability in the neighborhood of $\mathbf{\bar{x}}$ in the time- or frequency-domain. Small-signal analysis allows to investigate the dynamic properties of grid oscillatory phenomena in different frequency ranges, as reported in Table \ref{tab:static_dyn_linear_analysis}. Due to the nonlinearity of the IBR model, the small-signal properties of the IBRs can vary at different equilibrium points \cite{cecati_modeling_journal,yang_imped_datadriven}. \par
Large-signal dynamic stability analysis -- commonly denominated transient stability analysis -- is used to inspect nonlinear dynamics and instability phenomena (see Table \ref{tab:static_dyn_linear_analysis}) that are conceptually reconducted to the situation shown in Fig. \ref{fig:stability-ball}(c), with a stable equilibrium point with limited region of attraction. IBR transient instabilities imply that the system state is driven outside the region of attraction \cite{graz_regofattr}; such phenomena are often caused by the triggering of physical constraints (e.g. \cite{linbin_currlim,juancamilo_transient}), and/or by sluggish control response \cite{xiongfei_fastslow}. Transient stability is commonly analyzed through time-domain nonlinear IBR models or energy functions in the classic power system literature (see bottom of Fig. \ref{fig:stability-ball}(c)). \par 
In the context of IBRs, ensuring stability means avoiding \emph{all} three types of instability situations described in Table \ref{tab:static_dyn_linear_analysis} in a wide range of grid scenarios. Achieving this is highly challenging, since some IBR dynamics -- such as the synchronization -- affect multiple aspects of the stability in a contrasting way (e.g. control tunings that improve one aspect may deteriorate another one), as pointed out in Table \ref{tab:physic_constraints}. Furthermore, the coupling mechanism among different IBR control loops, IBR physical constraints (e.g. current and DC voltage limits) and the dynamics of the AC power grid are complex and multifold, and can lead to instability phenomena that remain elusive to traditional modeling and analysis tools \cite{kundur_book,timgreen_microgrid}. \par 
 These considerations motivate a profound rethinking of model-based stability analysis tools in a broader sense, to focus not only on one single aspect (e.g. small-signal stability), but rather investigating the \emph{interplay} between different stability aspects (e.g. static and dynamic, small- and large-signal), and on the interactions among diverse IBR dynamics and physical constraints during the instability mechanism.

\subsection{Nonlinear Modeling of Grid-Forming IBR}
\begin{figure}[t]
	\centering
	\includegraphics[width= \linewidth]{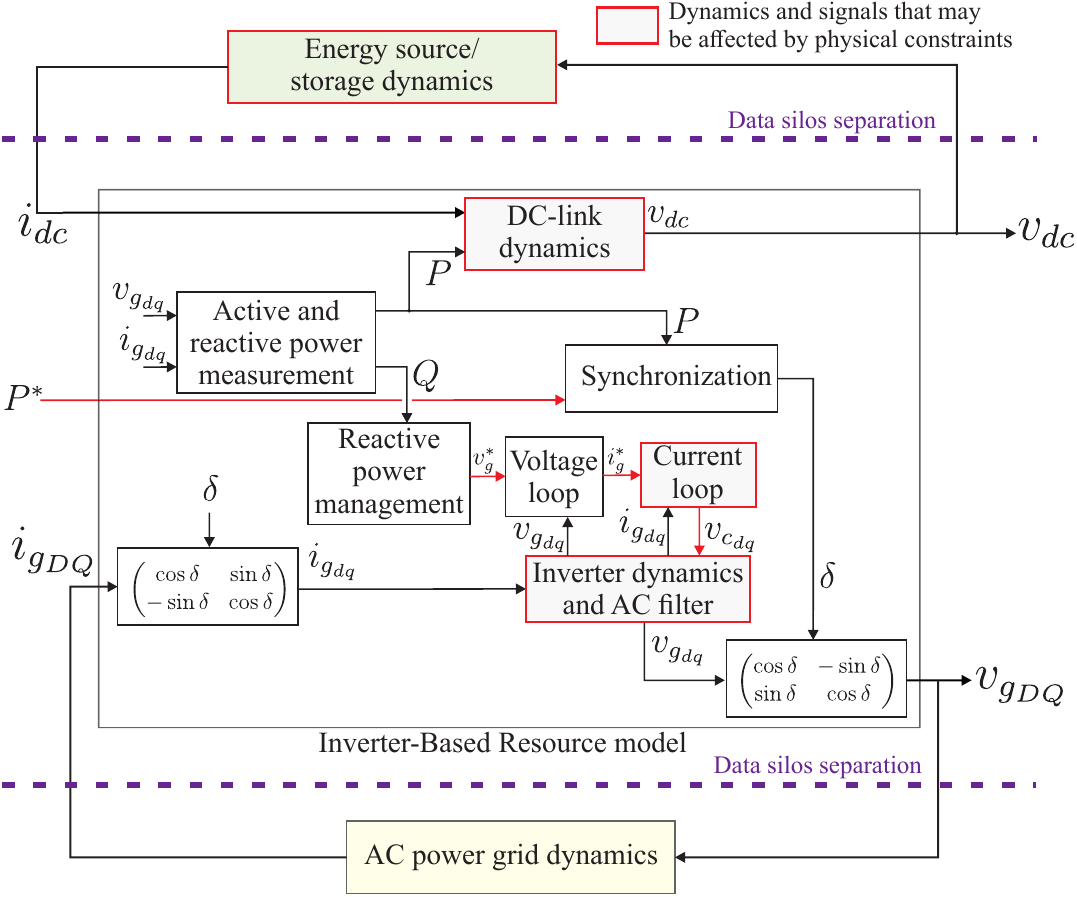}  
	\caption{Model structure including the different dynamics and their interrelations in a Grid Forming IBR model. }
	\label{fig:GrFRM_highorder_loops}
\end{figure}

To investigate the interplay between small- and large-signal aspects, a grid-forming IBR connected to a power grid is considered, without loss of generality for the presented methodologies. The corresponding model structure is illustrated in Fig. \ref{fig:GrFRM_highorder_loops}, where the different IBR dynamics described in Table \ref{tab:physic_constraints} are represented as blocks, each of them modeled through differential and/or algebraic equations \cite{cecati_modeling_journal}. The blocks and signals marked with a red color in Fig. \ref{fig:GrFRM_highorder_loops} are the ones that can be potentially affected by physical constraints. The reference current $i_g^*$, the reference output voltage $v_g^*$ and the reference inverter voltage $v_c$, for example, may be subject to variations during current limiting mode \cite{currentlim_overview}. The considered IBR model is aimed to study stability phenomena in a timescale below 1 kHz, which are the ones that can be effectively addressed through proper IBR control system design. The electromagnetic transient phenomena above 1 kHz (e.g. semiconductor switching) are neglected. \par 
The synchronization block, assuming a power synchronization for the angle $\delta$, is modeled as \cite{roberto_overview_gridforming}:
\begin{align}
    \begin{cases}
        \dot{\delta} &= m_p (P^* - P_{flt}) \\
        \dot{P}_{flt} &= \dfrac{1}{T_p} (- P_{flt} + P)
    \end{cases}
    \label{eqn:first-order_droop}
\end{align}
where $P$ is the injected active power, $P^*$ the reference power, $P_{flt}$ the filtered active power with time constant $T_p$, and $m_p$ the frequency droop gain; \eqref{eqn:first-order_droop} is equivalent to a Virtual Synchronous Machine implementation, where the active power low-pass filter emulates inertia \cite{salvatore_equivalence}. The power measurement block uses the well-known formulas in the $dq$-frame \cite{marcos_book}.
%
The DC-link dynamics is modeled through the power balance equation (neglecting losses), being $C_{dc}$ the DC capacitor \cite{cecati_modeling_journal}:
\begin{align}
    \dot{v}_{dc} =  \frac{1}{C_{dc}} \left( -\frac{3}{2}\frac{ P }{{v}_{dc}} + i_{dc} \right)
	\label{eqn:v_dc}
\end{align}
The voltage and current loops in Fig. \ref{fig:GrFRM_highorder_loops} are generally modeled through Proportional, Proportional Integral (PI), Proportional Resonant (PR) or Virtual Impedance (VI) loops, depending on their implementation. More details on their modeling can be found in \cite{marcos_book}. 
The AC magnitude set-point block in Fig. \ref{fig:GrFRM_highorder_loops} is assumed to be a droop control in the form:
\begin{align}
    \begin{cases}
        v_g^* &= V_n - n_q \left( Q^* - Q_{flt} \right)  \\
        \dot{Q}_{flt} &= \dfrac{1}{T_q} (- Q_{flt} + Q)
    \end{cases}
    \label{eqn:reactive_pow_droop}
\end{align}
where $V_n$ is the nominal voltage, $Q_{flt}$ is the low-pass filtered reactive power with time constant $T_q$, $Q^*$ is the reference reactive power \cite{roberto_overview_gridforming}.
The coordinate change between local $dq$ reference frame (internal to the IBR) and global $DQ$ reference frame (common to all IBRs of the grid) is realized through a rotation matrix, as in Fig. \ref{fig:GrFRM_highorder_loops} \cite{cecati_modeling_journal, timgreen_microgrid}. The $dq$ subscript in local frame is omitted, when unnecessary, for compactness of notation (e.g. $v_g =v_{g_{dq}}$).
The Inverter Dynamics and AC Filter block in Fig. \ref{fig:GrFRM_highorder_loops} includes in most of the cases only the L, LC, or LCL filter model, derived in \cite{marcos_book}, however recent researches have also taken into account thermal dynamics of semiconductors to study overcurrent capability (see \cite{cecati_transient_MMC}). 
The energy source/storage dynamics in Fig. \ref{fig:GrFRM_highorder_loops} is modeled simply as an ideal DC current source $i_{dc}$ regulating the DC voltage through a PI controller, though recent publications are realizing more detailed models of different resources (e.g. electrolyzer, wind turbine, flywheel) to analyze their impact on the IBR stability during grid contingencies \cite{flywheel_grfrm,electrolyzer_grfrm,PMSM_gridforming}.
The AC grid dynamics in Fig. \ref{fig:GrFRM_highorder_loops} is modeled, in its most simplified form, through a Thevenin equivalent with an RL impedance, with $i_g$ being the IBR current, $v_g$ the IBR voltage and $e$ the grid voltage. More general and informative models of power grids can be realized by leveraging graph theory \cite{graph_theory_florian}.
\begin{figure}[t]
	\centering
	\includegraphics[width=0.65 \linewidth]{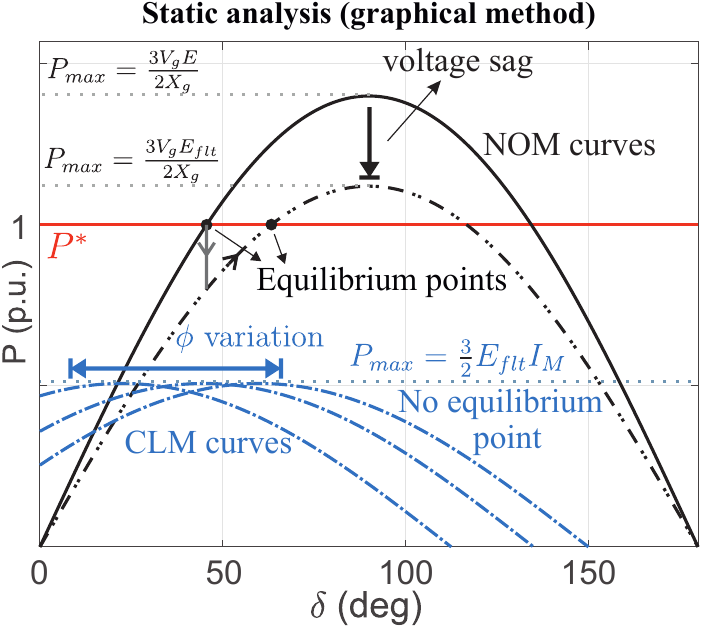}  
	\caption{ Graphical static large-signal analysis through $P$-$\delta$ curves. }
	\label{fig:Pdelta_curve}
\end{figure}
\subsection{Static large-signal stability analysis} \label{subsec:static_large-signal}

The simplest formulation of the static equation for IBR large-signal analyses is the well-known $P$-$\delta$ relationship, graphically represented in Fig. \ref{fig:Pdelta_curve} \cite{kundur_book}. $E$ is the grid voltage magnitude, $V_g = \| v_g \|$, and $X_g$ is the grid reactance. The intersection between the set-point $P^*$ (in red) and the Normal Operating Mode (NOM) $P$-$\delta$ curve (black solid line) defines the equilibrium points of the IBR state $\delta$ in \eqref{eqn:first-order_droop}. Voltage sags conditions are studied by modifying the value of $E$ from nominal to the faulty value $E_{flt}$, which reshape the $P$-$\delta$ curve as in Fig. \ref{fig:Pdelta_curve} (black dashed line). During large faults with Current Limitation Mode (CLM) activation -- assumed to be a magnitude current limiter where the IBR injects its maximum current $I_M \angle \phi$ in compliance with the grid regulations \cite{linbin_currlim, gabsuseo_overview_currlim,xavier_transint_rokrok} -- the $P$-$\delta$ curve is further reshaped as in the blue dashed curves of Fig. \ref{fig:Pdelta_curve}. The reduction of the maximum power transfer capability during faults is evident both in NOM (black dashed line) and even more in CLM (blue dashed line), and is a well-known cause of large-signal instability, for which several mitigation strategies have been proposed \cite{gabsuseo_trans_fictious,cross-forming}. \par
Despite its practicality, conventional $P$-$\delta$ graphical methods are only valid under the assumptions of purely inductive grid impedance, single-machine-infinite-bus scenario, constant voltage $V_g = 1$ p.u. and full decoupling among $P$-$\delta$ and $Q$-$V$ relationships \cite{kundur_book}, that may be unrealistic in real-world conditions. Furthermore, once the equilibrium point is computed, the $P$-$\delta$ curve provides limited information on the small-signal stability of the IBR at the computed equilibrium point. \par
\begin{figure}[t!]
	\centering
	\begin{subfigure}{0.49\textwidth}
		\centering
		\includegraphics[width= \linewidth]{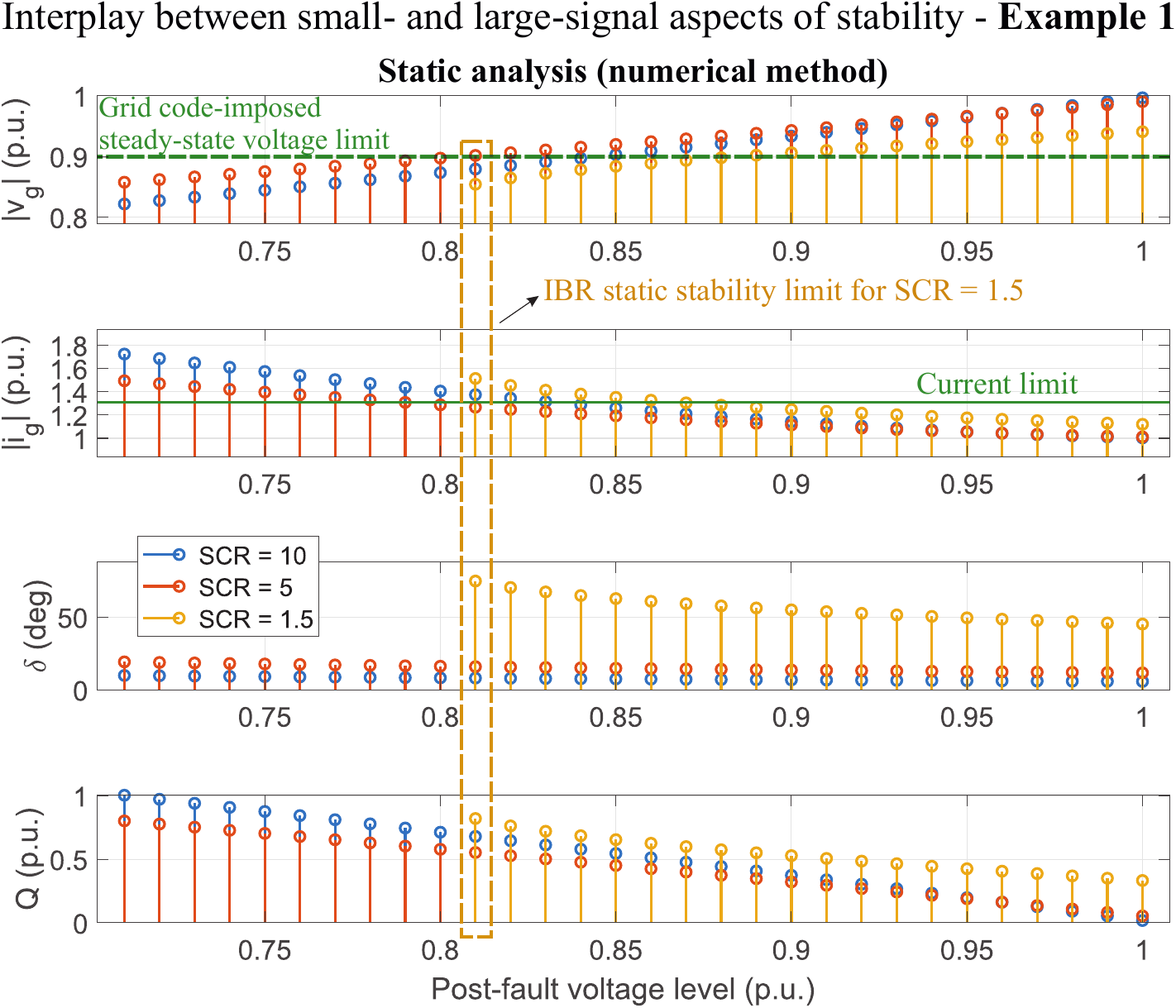}    
		\caption{ }
        \label{fig:static_iter}
	\end{subfigure}
	\begin{subfigure}{0.41\textwidth}
		\centering
        \includegraphics[width= \linewidth]{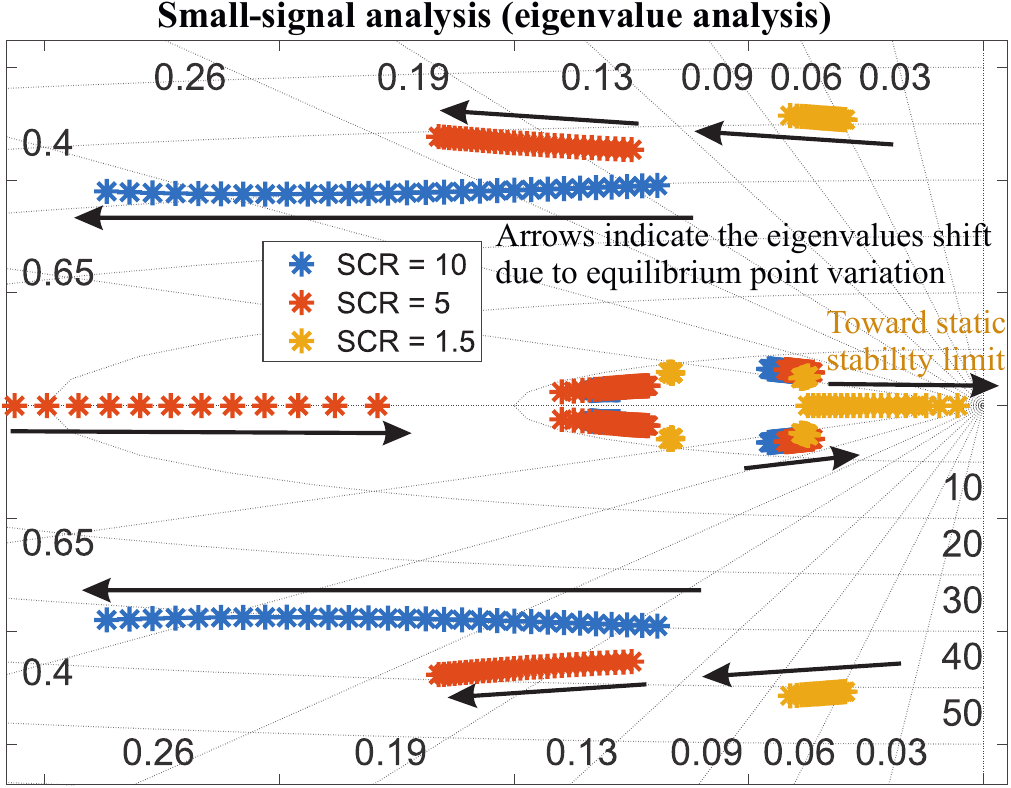} 
		\caption{ }
        \label{fig:eig_static_valid}
	\end{subfigure}
    \begin{subfigure}{0.49\textwidth}
		\centering
        \includegraphics[width=\linewidth]{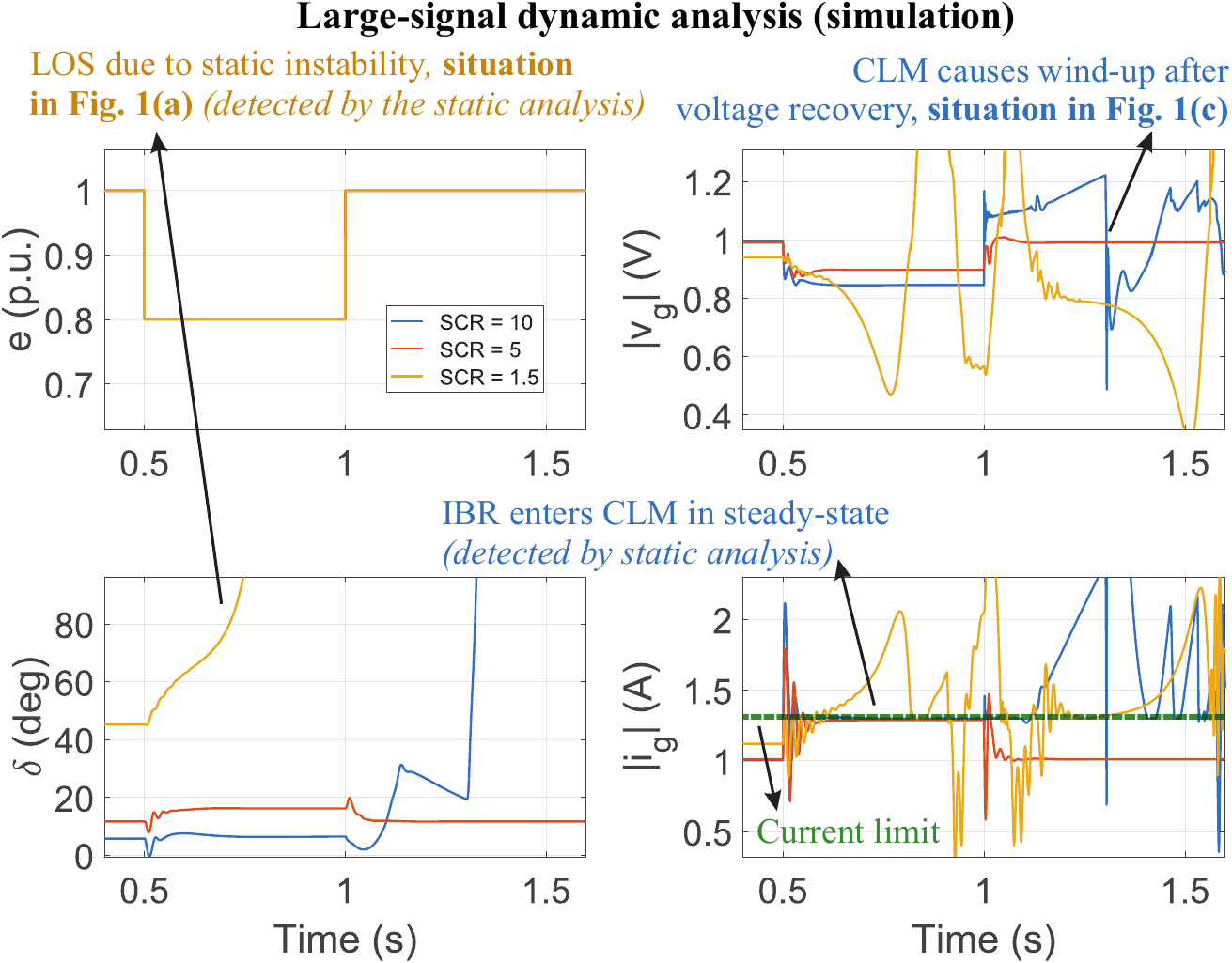} 
		\caption{ }
        \label{fig:simul_SCR}
	\end{subfigure}
	\caption{ Capability of different analytic tools to detect and study various instability phenomena caused by a voltage sag. (a) Static analysis by using numerical method. (b) Small-signal eigenvalue analysis at the computed equilibria. (c) Time-domain simulation. }
    \label{fig:interplay1}
\end{figure}

 
To overcome these limitations, more detailed $dq$-frame IBR models combined with \emph{numerical solution} of the static equation can be used \cite{cecati_modeling_journal}. The latter methodology, traditionally employed in machine-dominated phasor-based models \cite{costas_book,fmilano_opensource}, allows to capture the interplay between static and small-signal phenomena by linearizing the IBR model around the computed equilibrium point $\mathbf{\bar{x}}$. 
In Fig. \ref{fig:static_iter}, numerical static analyses is applied to the considered grid forming IBR model in Fig. \ref{fig:GrFRM_highorder_loops} under different depth of voltage sags and with three different values of Short Circuit Ratio (SCR).

The non-existence of equilibrium point, occurring with SCR of 1.5 for voltage sags to a value lower than 0.8 in Fig. \ref{fig:static_iter}, results in the \emph{unsolvability} of the static algebraic equation, and consequently the \emph{non-convergence} of the numerical algorithm for the post-fault static analysis \cite{andreas_voltstab,bifurcation_steadystate_loss}. \par 
From the obtained results in Fig. \ref{fig:static_iter}, the following conclusions on the steady-state operation of the IBR can be drawn:
\begin{itemize}
    \item The steady-state voltage magnitude $\|v_g\|=V_g$ decreases during faults due to the reactive power management loop in \eqref{eqn:reactive_pow_droop}, depending on its gain $n_q$, usually designed according to the grid code.
    \item The steady-state voltage magnitude $V_g$ varies both depending on the depth of the voltage sag and the SCR. Its reduction is a further reason for the shrinking of the power transfer capability.
    \item The activation threshold of the CLM in steady-state depends on the SCR, which influences the required amount of active/reactive currents to be injected during the faults to support the voltage.
    \item Static instability occurs for $E<0.8$ with SCR = 1.5 in Fig. \ref{fig:static_iter}, independently on the CLM. 
    \item When CLM is triggered in steady-state (e.g. $E<0.87$ with SCR = 1.5), it can cause CLM-induced static instability \cite{linbin_currlim}, due to the reduction of $P$ that makes \eqref{eqn:first-order_droop} unsolvable in static condition. Control strategies to properly adjust the reference power $P^*$ have been proposed to tackle this issue \cite{gabsuseo_trans_fictious,currentlim_overview}.
    \item In all the considered cases, CLM-induced static instability occurs with higher post-fault voltage levels with respect to the IBR static instability limits.
\end{itemize}
Most of the considered details, together with the interplay with small-signal aspects, are neglected by conventional $P$-$\delta$ graphic method, whose results can thus suffer from inaccuracy and limited applicability.
An important consideration for the application of numerical algorithms is their initialization with a flat start $\mathbf{{x}}_0$ not too far away from the solution $\mathbf{\bar{x}}$, to avoid non-convergence due to numerical issues, by using the procedure described in \cite{cecati_LTP_paper}. 
\begin{figure}[t!]
	\centering
	\begin{subfigure}{0.47\textwidth}
		\centering
		\includegraphics[width= \linewidth]{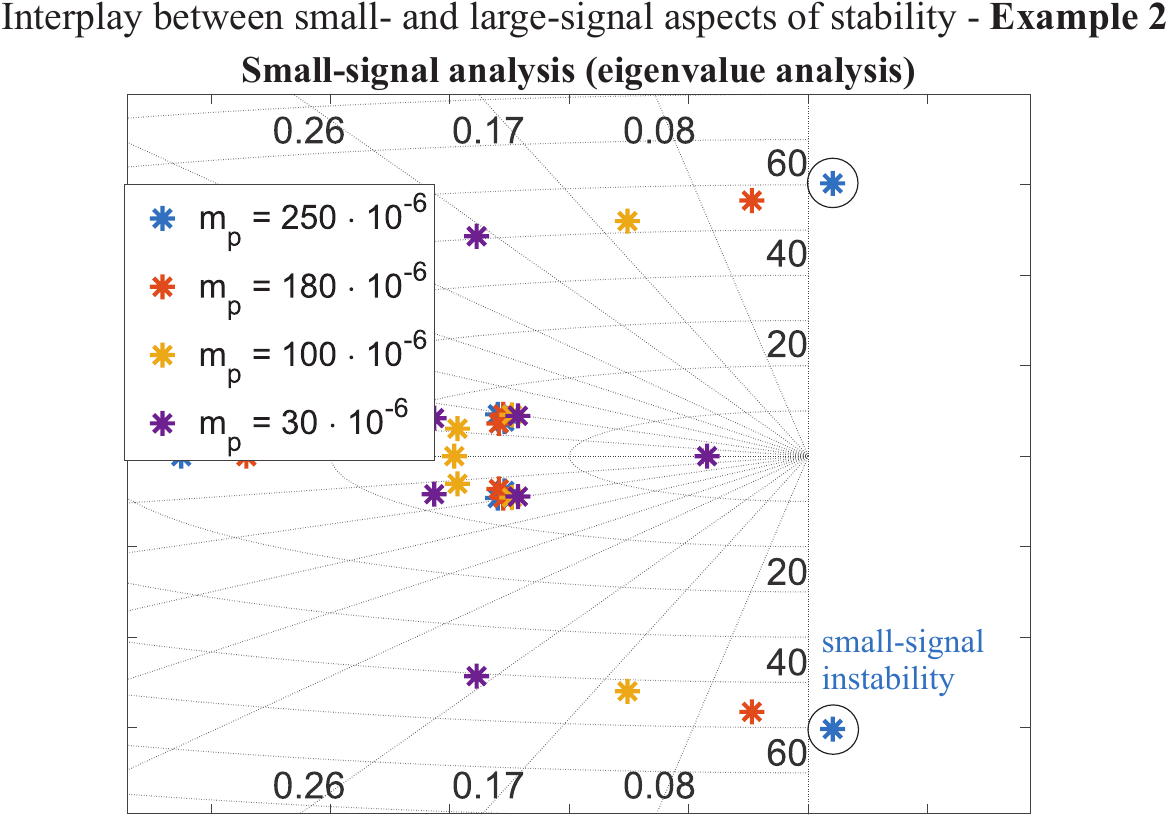}   
		\caption{ }
        \label{fig:eig_mp}
	\end{subfigure}
	\begin{subfigure}{0.499 \textwidth}
		\centering
        \includegraphics[width=\linewidth]{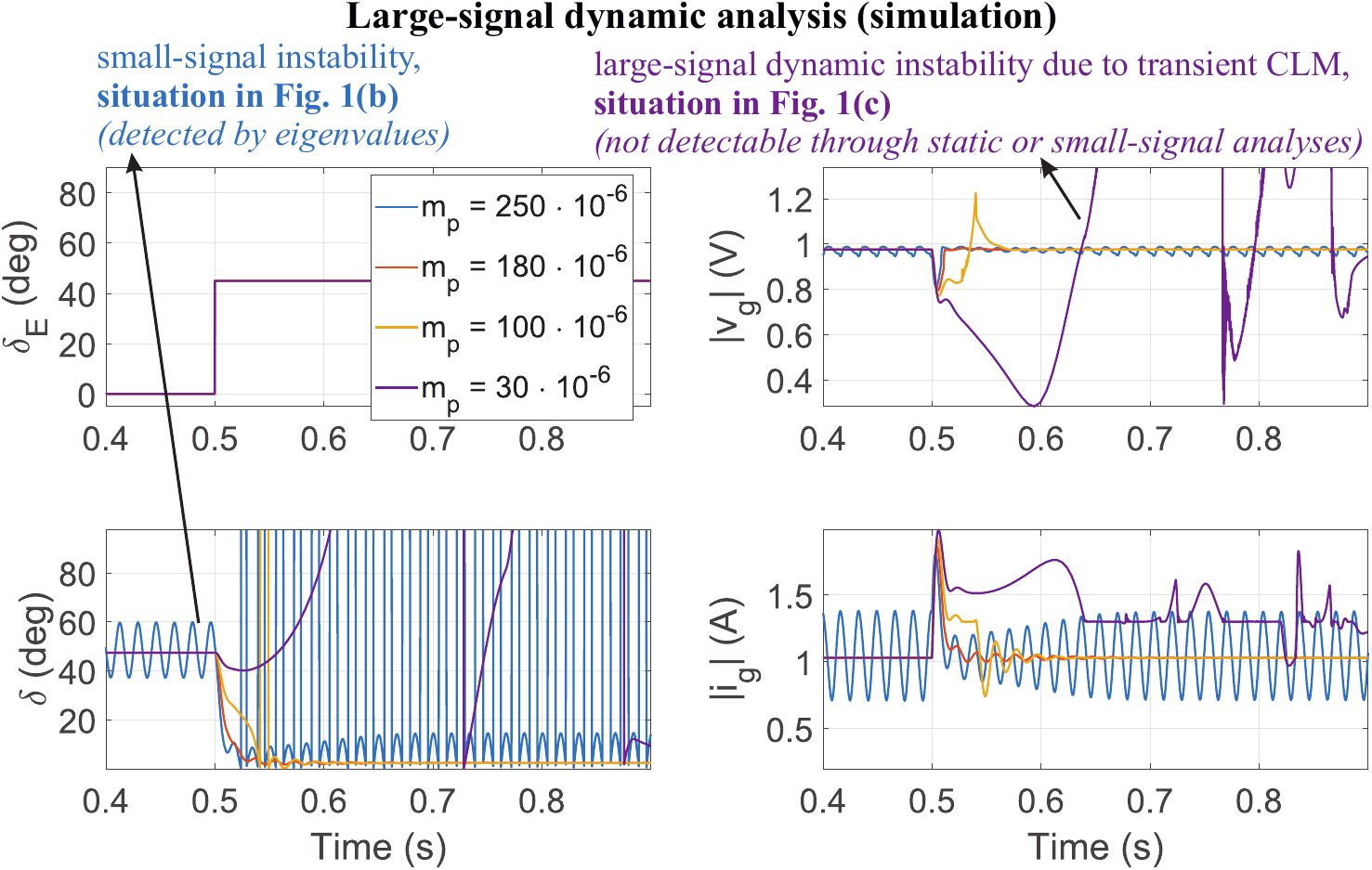} 
		\caption{ }
        \label{fig:mp_LOS_n}
	\end{subfigure}
	\caption{ Capability of different analytic tools to detect and study various instability phenomena caused by a phase jump. (a) Small-signal eigenvalue analysis. (b)  Time-domain simulation. The static analysis is omitted, since $m_p$ does not affect the equilibrium point.  }
    \label{fig:interplay2}
\end{figure}

\subsection{Small-signal stability analysis} \label{subsec:small_signal} 
Small-signal instability in IBR-based grids includes diverse oscillatory phenomena at different frequency ranges (reported in Table \ref{tab:static_dyn_linear_analysis}) that have been among the driving causes of recent blackouts \cite{spanish_blackout_finalreport}. Small-signal stability analysis is carried out on the IBR model linearized around $\bar{\mathbf{x}}$, computed through the static analysis in subsection \ref{subsec:static_large-signal}, if it exists. The analyses can be performed using either time-domain (e.g. eigenvalue analysis) or frequency-domain methods (e.g. Nyquist criterion) \cite{xiongfei_harmstab_first,timgreen_microgrid}. Robust stability tools \cite{cecati_interop,roberto_synchro_parall}, passivity \cite{Harnefors_Passivity_2016}, and small-gain/small-phase theorems \cite{linbin_gainphase,mattavelli_specification} have been also proposed for assessing small-signal stability and its margins.
\subsubsection{\textbf{Oscillation Damping Capability Evaluation}}
Modern grid requirements set stringent requisites for the oscillation damping capability of IBRs, often expressed in terms of minimum damping ratio (e.g. ENTSOE GFM requirements \cite{entsoe_gfm_phase2_2025}). Eigenvalue analysis is the most straightforward tool to precisely characterize the damping ratio of the different IBR oscillations modes \cite{book_phasemargin}.
Grid parameters such as the Short Circuit Ratio (SCR) have a widely-recognized effect on the IBR stability \cite{behnam_delay_conf,behnam_delay_openjournal,behnam_delay_shanhe,lingling_PV_BESS}. The mechanism of influence is, however, \emph{two-fold}. Firstly,  SCR variations are directly reflected on the IBR eigenvalues and small-signal properties (see Fig. \ref{fig:eig_static_valid}). Secondly, SCR variations affect the static IBR equilibrium point (see Fig. \ref{fig:static_iter}), which in turn influences the IBR small-signal properties, due to the nonlinearity of the IBR model. The latter indirect effect is a clear example of interplay between static and small-signal stability aspects. \par 
In Fig. \ref{fig:eig_static_valid}, the shift of the eigenvalues with decreasing grid voltage levels (same levels considered in the $x$-axis of Fig. \ref{fig:static_iter}) is indicated by the arrows for the three SCR values. 
The IBR oscillations damping capability (related to the eigenvalues' damping ratio) varies both by the grid voltage level and the SCR, making the compliance with grid technical requirements more difficult \cite{ngeso_gfm_2025,entsoe_gfm_phase2_2025}.
Furthermore, a purely real eigenvalue in Fig. \ref{fig:eig_static_valid} moves toward the origin when the grid voltage level decreases. This eigenvalue is not related to small-signal oscillatory phenomena, but rather to the fact that the Jacobian matrix of the IBR nonlinear model becomes near to singular when approaching the static stability limit, where a saddle-node bifurcation occurs \cite{bifurcation_gridforming,bifurcation_hyperplanes,bifurcation_steadystate_loss,chi_ktse_bifurcation,chi_ktse_bifurcation2_rectifier}. \par 
Another eigenvalue analysis is shown in Fig. \ref{fig:eig_mp} with respect to the frequency droop gain $m_p$. This analysis detects a case of small-signal instability for $m_p = 250 \cdot 10^{-6}$, and will be discussed more in depth in subsection \ref{subsec:dynamic-large-signal}.

\subsubsection{\textbf{Impedance-based small-signal stability analysis}} \label{subsec:small-signal_freqdom}

\begin{figure}[t]
	\centering
	\includegraphics[width= \linewidth]{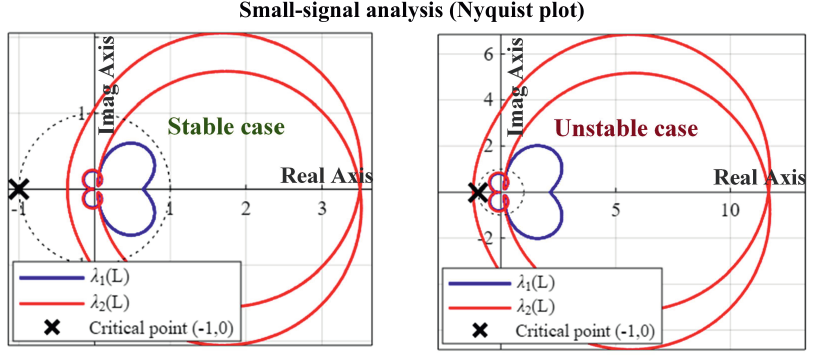}  
	\caption{ Nyquist-based stability evaluation for MIMO IBR models, considering the eigenvalues of the return-ratio matrix.}
	\label{fig:Imp}
\end{figure}

\begin{figure*}[t]
	\centering
	\includegraphics[width=0.99 \linewidth]{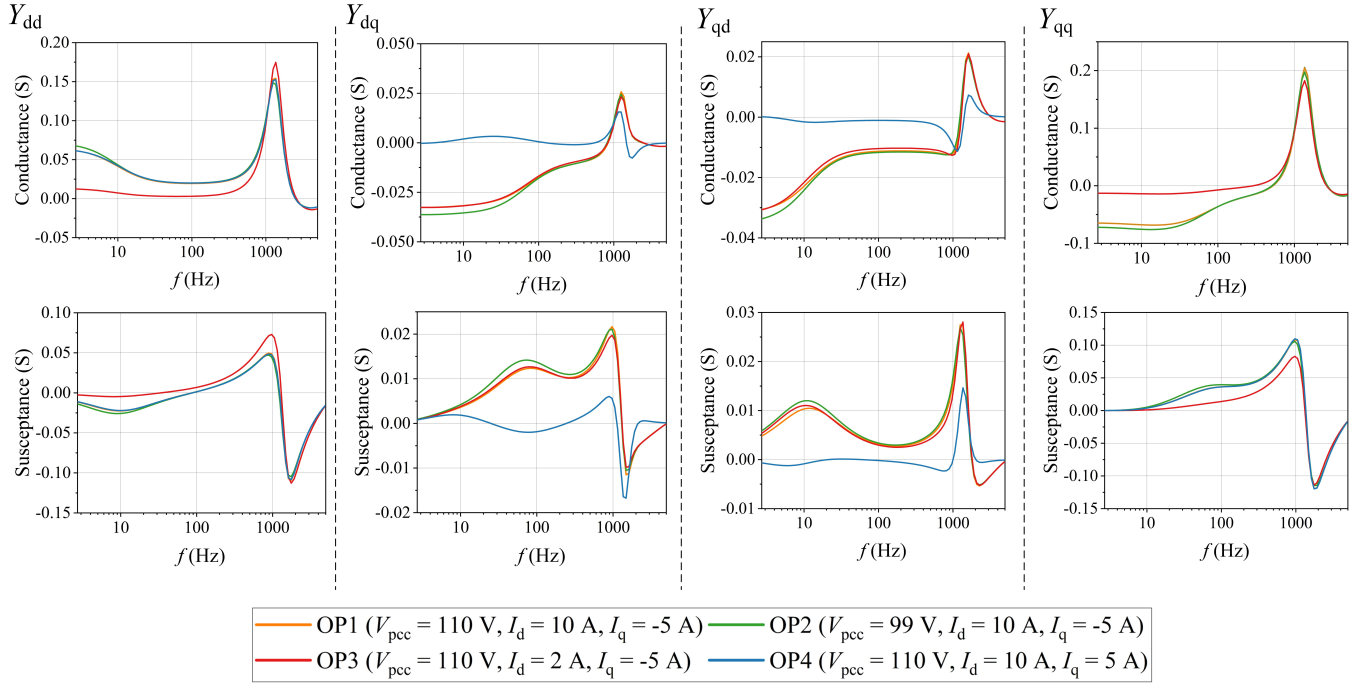}  
	\caption{ Impedances of a GFL-IBR at four different operating points (OPs). The configuration of the IBR can be found in \cite{yang_imped_datadriven}. $V_{\text{pcc}}$ represents the PCC voltage, while $I_{\text{d}}$ and $I_{\text{q}}$ denote the current at different axes. As shown, the impedance profile exhibits notable operating-point dependence, most prominently at low frequencies. Each channel of the impedance profile (expressed in admittance form $Y_{\text{dd}}$, $Y_{\text{dq}}$, $Y_{\text{qd}}$, and $Y_{\text{qq}}$) has a conductance component and a susceptance component.}
	\label{fig:Overall1}
\end{figure*}

Impedance-based stability analysis provides a systematic framework for investigating the dynamic interactions between the IBRs and the rest of the power grid. 
Small-signal impedance (or admittance) models characterize IBR dynamics in the frequency-domain at their AC terminals (i.e. $v_{g_{DQ}}$ and $i_{g_{DQ}}$ in Fig. \ref{fig:GrFRM_highorder_loops}).  
Normally, the IBR impedance is modeled as a multi-input multi-output (MIMO) transfer function matrix to capture the $dq$-frame cross-coupled dynamics, and studied through the generalized Nyquist criterion -- i.e. analyzing the eigenvalues of the return-ratio matrix -- as shown in Fig. \ref{fig:Imp}. 
In accordance with the results in Fig. \ref{fig:eig_static_valid}, the variation of IBR operating point significantly affect its small-signal impedance, specially in the low-frequency range \cite{Wen_Analysis}, as confirmed by Fig. \ref{fig:Overall1}.
%
%
\subsubsection{\textbf{Decentralized small-signal stability criteria}}
Passivity-based stability criteria are increasingly incorporated in modern grid codes (e.g. ENTSOE \cite{entsoe_gfm_phase2_2025}), and offers a promising decentralized method for ensuring stability of IBR-dominated grids, grounded in the principle that the interconnection of passive subsystems is inherently stable \cite{Harnefors_Passivity_2016}. In the frequency-domain, an IBR is considered passive if its impedance $Z(s)$ (or admittance $Y(s)$) has a non-negative real part -- which physically corresponds to a positive resistance -- across all frequencies, meaning $\text{Re}\{Z(j\omega)\} \geq 0$ or equivalently $\angle Z(j\omega) \in [-\pi/2  \ \ \pi/2]$ for all $\omega$ \cite{linbin_gainphase}. 
However, conventional passivity theory presents significant challenges at low frequencies, as the slow dynamics of synchronization and power control loops may result in non-passive regions \cite{lowfreq_passivity}. To mitigate this weakness, an extended frequency-domain passivity theory has been proposed \cite{Chen_Extended_2025}, which utilizes rotation matrices to provide extended stability conditions for MIMO systems. Another shortcoming of passivity-based approaches is conservativeness. While traditional passivity is a condition on the phase of $Z(j\omega)$, its proper combination with conditions on the gain of $Z(j\omega)$ and the use of frequency-split approaches sheds new light on novel decentralized stability certificates with low conservativeness \cite{linbin_gainphase, Chen_Unified_2025,verena_loop_shifting}. More recent advances include scaled relative graphs and Davis-Wielandt- (DW-) shells, discussed in depth in \cite{Leng_arXiv_2026}.
\begin{figure}[t]
	\centering
	\includegraphics[width= \linewidth]{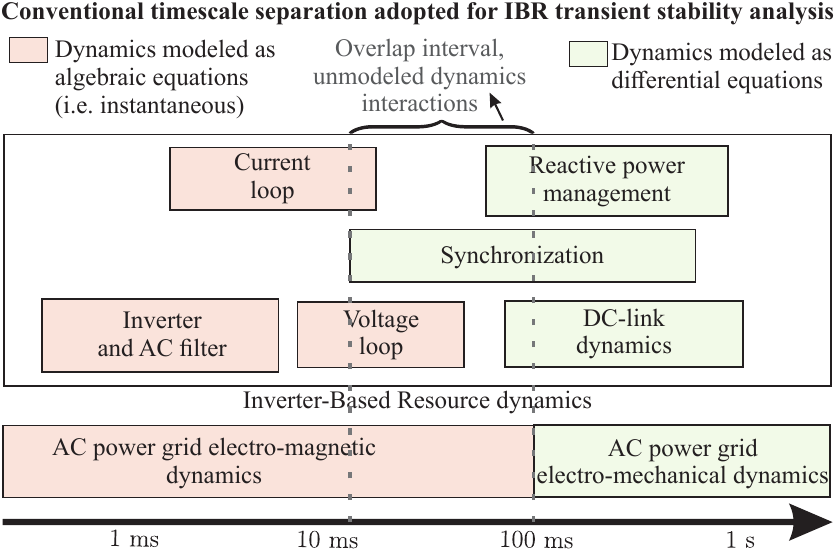} 
	\caption{ Timescale of different IBR dynamics, and their conventional modeling approach in transient stability studies.  }
	\label{fig:timescale_separation}
\end{figure}
\subsection{Dynamic large-signal stability analysis} \label{subsec:dynamic-large-signal}
%
The investigation of transient instability phenomena as the one in Fig. \ref{fig:stability-ball}(c), considering cross-coupling among diverse IBR dynamics and physical constraints, lacks an off-the-shelf solution, and leaves still nowadays many open questions. The existing methods fall into two main families: analytical methods, which do not need time-domain integration -- e.g. Equal Area Criterion, Lyapunov/Energy function -- and the time-domain integration methods -- e.g. phase portraits, simulation -- used in this subsection to show the interplay between small- and large-signal stability aspects.
%
%
%
%
%
%
%
\subsubsection{\textbf{Analytical methodologies}} \label{subsub:numerical_static_analysis}
The Equal Area Criterion is a simple and well-established energy-based analytical method for transient stability analysis. Initially proposed for SGs \cite{kundur_book}, it can be applied to IBRs modeled through their $P$-$\delta$ relationship (see Fig. \ref{fig:Pdelta_curve}) \cite{hengwu_trans_pow,linbin_currlim}. Its applicability is, however, limited to single-machine-infinite-bus systems, and its validity for IBR stability analyses has been recently questioned \cite{brian_johnson_EAC_is_applicable}. \par 
Lyapunov/energy-based analyses trace their origins to SG-dominated grids, where they have been studied for particular power grid configurations, e.g. multi-machine grid with lossless network \cite{energy_func_Pai_book}. More recently, rigorous proofs for global stability of specific IBR control strategies have been derived \cite{dvoc_proof_xiuqian_decenstralized,dvoc_proof_xiuqian_decenstralized2,matching_control_florian, hybrid_angle_control,cross-forming}. The presence of non-smooth control dynamics, such as the current limitation, significantly complicate Lyapunov analyses. To tackle this issue, control barrier functions are currently under investigation for constraint-aware IBR control with stability warranties \cite{control_barrier_function_sijia}.  \par  
Most transient stability analyses make use of singular perturbation theory to separate the IBR and grid dynamics timescales into a faster subsystem (including subtransient dynamics, assumed instantaneous) and a slower subsystem (including transient and quasi-stationary dynamics), as shown in Fig. \ref{fig:timescale_separation} in red and green, respectively \cite{khalil_book,irina_singpert}.
A common assumption is that the typical phenomena responsible of transient instability (e.g. LOS, exceeding maximum power transfer capability) are mostly related to the dynamics represented in green in Fig. \ref{fig:timescale_separation} \cite{hengwu_trans_PLL,cross-forming}. However, these dynamics (e.g. synchronization and outer loops) often operate in a broader timescale, which partially overlaps with the electromagnetic dynamics of power network and the inner IBR control loops dynamics, that are often neglected in transient studies \cite{rethinking_currctrl,harm_stab_concept}. This timescale overlap reveals the inadequacy of quasi-stationary IBR and grid models for transient stability studies in IBR-based grids, and underlies the current debate on the necessity of more detailed models such as dynamic phasor, nonlinear state-space models as in \cite{cecati_modeling_journal}, or EMT models \cite{paolone2020fundamentals, harm_stab_concept, EMT_mahseredjian}. \par 
Lyapunov/energy-based methods for transient stability analysis are poorly scalable to larger and more general IBR-based grids, since energy function derivation is often highly model-specific and can suffer from the lack of a systematic methodology. In such cases, numerical methodologies such as Sum-Of-Square allow to build Lyapunov functions by means of computational techniques, and have been successfully applied to small-scale power grids and IBRs \cite{mike_sum_of_square}. For more complex systems, learning Lyapunov functions by means of neural networks opens up to a higher degree of scalability and generalizability, and granted promising results in recent studies \cite{lyap_neuralnetwork_florian}.
%
%
%
%
%
%
%
\subsubsection{\textbf{Time-domain simulation}}
Time-domain simulations benefit from a high scalability, and are used to study highly complex large-signal phenomena that can not be investigated through analytical methods. They can assess the stability only for a limited number of initial conditions and parameter settings, their results are often hard to interpret, and they can not give rigorous stability guarantees such as Lyapunov-based analytical methods. To extend stability assessment to a wider range of scenarios, iterative simulations combined with search algorithms and machine learning to estimate the transient stability margins are under investigation \cite{juancamilo_transient,panos_transient_machinelearn,david_hill_SVM}. With large-scale grids, however, the high computational burden of simulation (specially with EMT models) calls into question the feasibility of such kinds of approaches. To tackle this issue, physics-informed neural network-based accelerators are currently under investigation to accelerate simulation executions while retaining trustworthiness \cite{karampinis2027neural}. \par 
In this subsection, time-domain integration is applied to full-order IBR nonlinear models in Fig. \ref{fig:GrFRM_highorder_loops} to validate the analyses in Figs. \ref{fig:static_iter}, \ref{fig:eig_static_valid} and \ref{fig:eig_mp}, and to further emphasize the interplay between small- and large-signal aspects of the IBR stability. \par 
Fig. \ref{fig:simul_SCR} depicts the simulation of a voltage sag of 0.8 p.u. with different SCR values, to complement the static analysis in Fig. \ref{fig:static_iter} and the small-signal analysis in Fig. \ref{fig:eig_static_valid}. For SCR = 10, the IBR injects high reactive current to control the voltage, and reaches the overcurrent during the 0.8 p.u. voltage sag, confirming the results in Fig. \ref{fig:static_iter}. Since the fault duration is relatively short and the overcurrent is not severe, the IBR remains initially stable. However, after the fault clearing, the IBR experiences instability due to the windup induced by the previous CLM. This kind of instability can be mitigated by applying anti-windup strategies both to the current loop and the synchronization integrator \cite{gabsuseo_trans_fictious,gabsuseo_currlim_smallsign}. \par 
In the case of SCR=1.5 with a 0.8 p.u. voltage sag, the IBR equilibrium point does not exist (see Fig. \ref{fig:static_iter}), and consequently the IBR experiences LOS as confirmed by Fig. \ref{fig:simul_SCR}. 
The case with SCR=5 is the only stable case in Fig. \ref{fig:simul_SCR}: no CLM occurs in steady-state (see Fig. \ref{fig:static_iter}), the equilibrium point exists with low power angle $\delta \approx 20$deg (see Fig. \ref{fig:static_iter}), and is small-signal stable with sufficient damping ratio (see Fig. \ref{fig:eig_static_valid}).  \par 
%
%
Fig. \ref{fig:mp_LOS_n} shows the simulation of a phase jump with different tuning of the synchronization loop in \eqref{eqn:first-order_droop}. For $m_p = 250 \cdot 10^{-6}$, the IBR experiences small-signal instability (i.e. sustained oscillations), confirming the analysis in Fig. \ref{fig:eig_mp}. The instability observed for $m_p = 30 \cdot 10^{-6}$ in Fig. \ref{fig:mp_LOS_n} is a dynamic large-signal instability (situation in Fig. \ref{fig:stability-ball}(c)), due to the large amplitude of the current response triggering CLM. 
High damping around the post-fault equilibrium point, which can be approximately quantified by the eigenvalues in Fig. \ref{fig:eig_mp}, has been proven by many papers to be beneficial for transient stability (see Fig. \ref{fig:stability-ball}(c)) \cite{hengwu_trans_PLL,transstab_nonlindamping,hengwu_trans_pow}. Damping ratio, however, describes only decay rate of oscillations, but gives no information on their \emph{amplitude}, which is exactly the driving cause of exiting the region of attraction in Fig. \ref{fig:mp_LOS_n} \cite{book_phasemargin}. This type of instability happening during phase jumps, related to triggering converter physical constraints (e.g. current limits, DC or AC voltage limits), is particularly difficult to analyze and mitigate. The lack of a general agreement on an analytical criterion for this type of instability translates into a difficulty in defining interpretable grid requirements for IBRs. Indeed, neither grid requirements on critical damping ratio nor on passive response are directly addressing this kind of phenomenon -- despite being overall beneficial for transient stability.

\subsection{Extension to multi-IBR systems}

The interplay between small- and large-signal dynamics exists -- and is even more relevant -- in multi-IBR power grids \cite{cecati_interop}. Its analysis requires a nonlinear model of the IBR-dominated grid, to apply the procedure at the bottom of Fig. \ref{fig:stability-ball}.
Component Connection Method (CCM), proposed in \cite{yanbo_CCM}, allows to interconnect linear IBR models solely for small-signal analyses.
To analyze the interplay of small- and large-signal stability aspects, a generalization of CCM for building nonlinear state-space models of power grids is necessary, as proposed in \cite{cecati_modeling_journal} and available in the repository \cite{repository_cecati_nonlinmod}. \par 
The static analysis in such multi-IBR grid translates into a generalized power flow analysis, to be solved with the numerical methods discussed in subsection \ref{subsec:static_large-signal}, detecting both bus voltage violations and also non-existence of equilibrium points \cite{cecati_modeling_journal,cecati_interop,behnam_delay_openjournal}.
Small-signal analysis in multi-IBR systems can be carried out both in time- and frequency-domain, where tools such as participation factor and sensitivity analysis are commonly used to detect which IBR is majorly responsible of generating undamped oscillations \cite{freqdom_participation,greybox_timgreen,yicheng_interoperab}. 
Large-signal stability warranties for multi-IBR systems have been derived by using decentralized stability conditions in \cite{dvoc_proof_xiuqian_decenstralized,dvoc_proof_xiuqian_decenstralized2}, yet they are valid only under specific assumptions and are not scalable to arbitrary grids with heterogeneous IBRs. 
In such scenarios where no analytical proof can be derived, simulation-based methods discussed in subsection \ref{subsec:dynamic-large-signal} are first-choice tools for large-signal dynamic studies, albeit computationally heavy for large networks \cite{panos_probabilistic_trasstab,panos_transient_machinelearn}.

\begin{figure}[t]
	\centering
	\includegraphics[width= \linewidth]{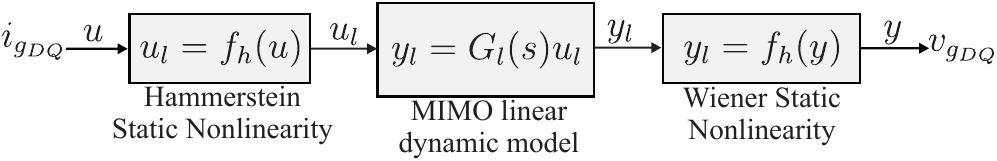}  
	\caption{ An example of Hammerstein-Wiener model parametrization for power converters.}
	\label{fig:hammerst_wiener}
\end{figure}
\begin{figure}[t]
	\centering
	\includegraphics[width=1 \linewidth]{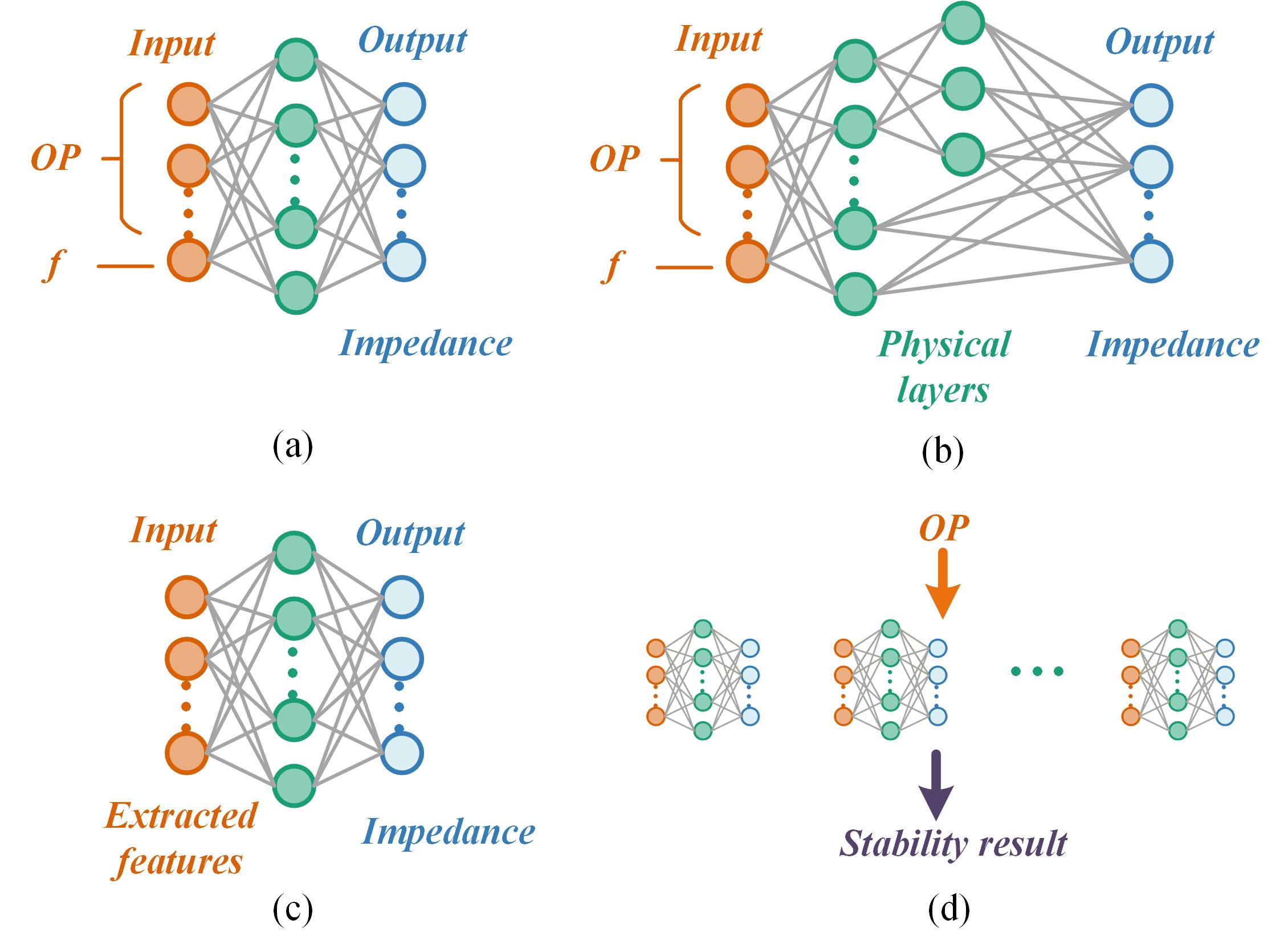}  
	\caption{ Neural-network-based stability analysis for IBRs. (a) The operating point (OP) is taken as the input. (b) Physics is embedded in the middle hidden layers. (c) The extracted features are taken as the input. (d) Ensembled neural networks for reliable stability analysis \cite{HumblotRenaux_UncertaintyAware_2025}.}
	\label{fig:NN1}
\end{figure}

\begin{figure}[t]
	\centering
	\includegraphics[width=0.75 \linewidth]{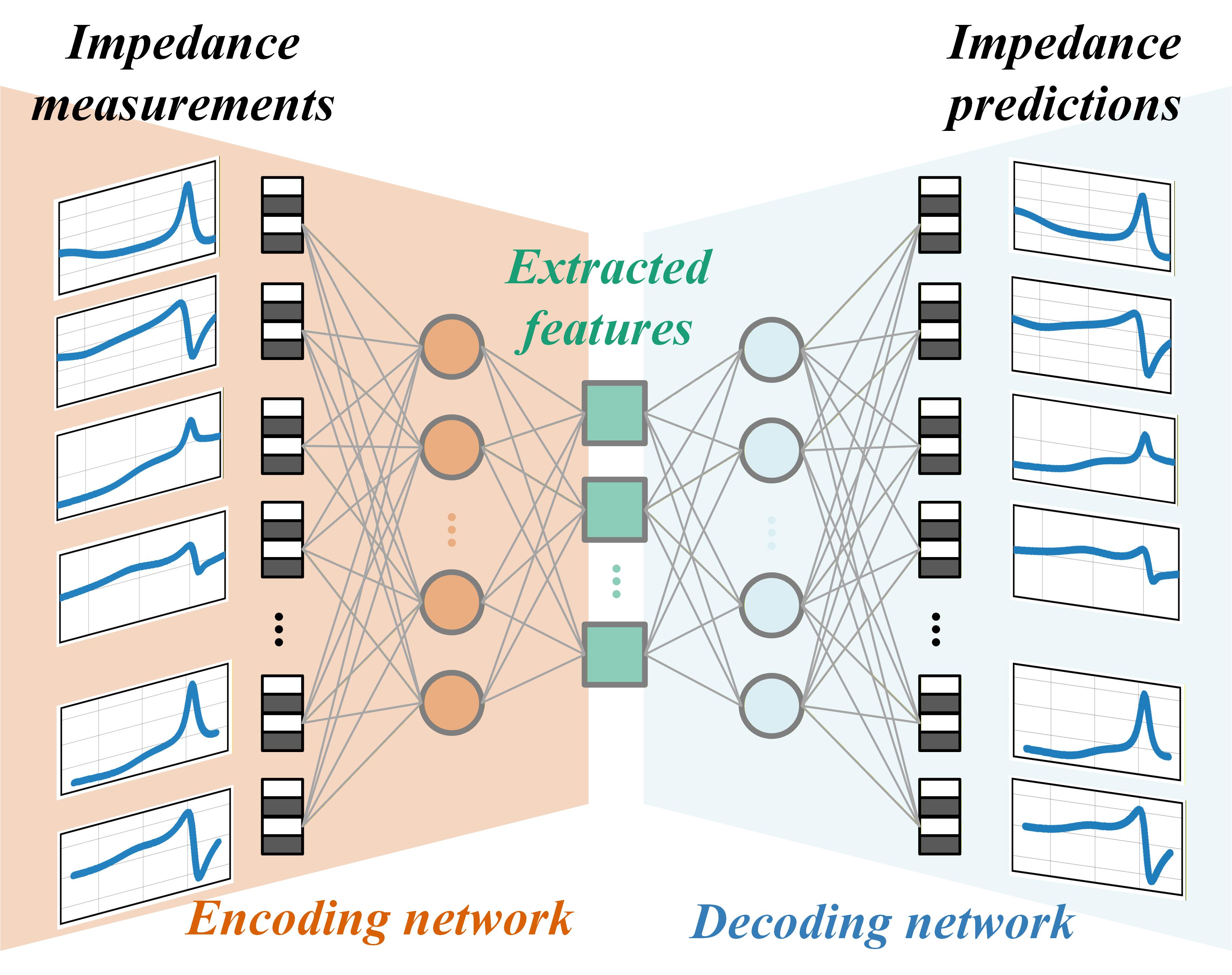}  
	\caption{ Feature-driven neural network architecture for IBR impedance modeling. The stacked autoencoder is used for feature extraction and impedance prediction \cite{yang_imped_datadriven}.}
	\label{fig:10a}
\end{figure}

\begin{figure*}[t]
	\centering
	\includegraphics[width=1 \linewidth]{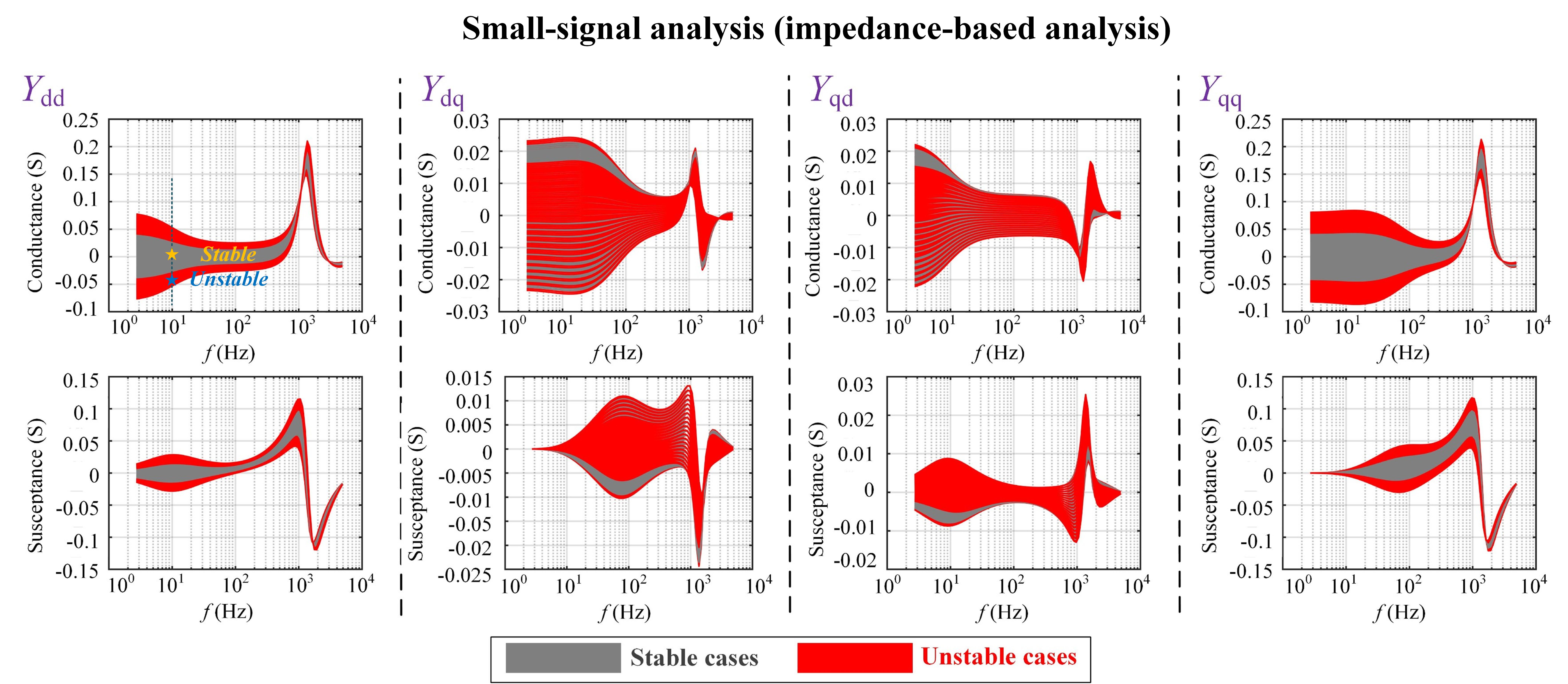}  
	\caption{ Predicted stability boundary in one channel, where curves in red indicate unstable cases and curves in grey represent stable cases \cite{yang_imped_datadriven}.}
	\label{fig:10b}
\end{figure*}
%

%
%
%
%
%
%
\section{Data-driven methodologies for stability analysis of IBR-dominated grids } \label{sec:data-driven}

Model-based methodologies discussed in Section \ref{sec:model-based} rely on the assumption of full knowledge of model details and parameters. In real-world scenarios, manufacturers and system operators do not share detailed information among each other, contributing to the creation of data silos -- i.e. private collection of data and model inaccessible to external stakeholders -- as illustrated in Fig. \ref{fig:GrFRM_highorder_loops}. The data silos separation between IBR and DC resource has been discussed in \cite{electrolyzer_grfrm}, and is kept out of the scope of this manuscript. The data silos separation between IBR and AC power grid, main topic of this section, has a dual implication. IBR models are not available to system operators, which encounter difficulties in analyzing power grid stability \cite{Zhang_ANN_2021}. On the other hands, IBR manufacturers have limited knowledge of the rest of the power grid, whose model is highly complex and time-varying, and encounter difficulties for the IBR control design.
These challenges motivate the adoption of data-driven methodologies which leverage measurements and operational data to assess and enhance stability in IBR-dominated grids. Considering the interplay between small- and large-signal dynamics is however even more challenging with data-driven methods than with model-based methods, and the available solutions will be discussed in this section. 

\subsection{Measurement and gray-box modeling for IBRs}
 To protect the sensitive intellectual property (IP) of the vendors \cite{Li_Neural}, data-driven models identified from simulation or experiment measurements have become the trend for industry standard, which can be provided to the external parties for stability analysis. In time-domain, the most direct model is an encrypted electromagnetic transient (EMT) model, sometimes provided by the vendor, which can be used to simulate the interactions of an IBR with the external grid, as done in Figs. \ref{fig:simul_SCR} and \ref{fig:mp_LOS_n} \cite{EMT_mahseredjian,EMT_mahseradjan_2}. While highly accurate, applying these models can be time-consuming by iterative calculation at each time step for different operating points. To address this limitation, a surrogate linear state-space model has recently been proposed, where the state-space matrix is transformed into diagonal form using an encrypted transformation matrix \cite{Smith_Black}. Thus, the eigenvalue information can be maintained, providing an accurate estimation of IBR dynamics responses. However, such state-space models remain dependent on the IBR operating point and introduce additional complexity. More critically, the physical meaning of the state variables is hidden, which can hinder their effectiveness for integrated IBR analysis.  \par 
To capture the large-signal dynamics of IBRs, gray-box models based on identification techniques have utilized Hammerstein-Wiener parametrization (conceptually depicted in Fig. \ref{fig:hammerst_wiener}) \cite{Buttner_Complex-Phase_2025}. Complex frequency and phase dynamics is captured through static nonlinear transformations, while phase angle and voltage magnitude nonlinearities are separated from a core linear subsystem. Accurate emulation of various control strategies can then be realized, offering a computationally efficient alternative to black-box EMT simulation models for large-scale stability assessments. \par 
In frequency-domain, impedance models based on measurements at different operating points can preserve the IP of different vendors while still allowing model identification to accurately capture system behavior for small-signal stability analysis, control and integration. To obtain sufficient impedance data, impedance measurements in a broad frequency range are always required at different operating points. Typically, in a three-phase IBR system represented in the $dq$-frame, cross-couplings between axes must also be considered. Therefore, two linearly independent perturbations are always injected into the $d$ and $q$ channels of the IBR system. Four equations can be obtained to calculate the different frequency-domain entries of the impedance matrix as shown in Fig. \ref{fig:Overall1} \cite{Gong_DQ}:
\begin{equation}
\begin{bmatrix}
U_{\text{d}1} & U_{\text{d}2}\\
U_{\text{q}1} & U_{\text{q}2}
\end{bmatrix} =
\begin{bmatrix}
Z_{\text{dd}} & Z_{\text{dq}}\\
Z_{\text{qd}} & Z_{\text{qq}}
\end{bmatrix}
\begin{bmatrix}
I_{\text{d}1} & I_{\text{d}2}\\
I_{\text{q}1} & I_{\text{q}2}
\end{bmatrix}.
\label{eqn:imped_mea}
\end{equation}
In (\ref{eqn:imped_mea}), $U_{\text{d}1}$ and $U_{\text{q}1}$ denote the $d$-axis and $q$-axis voltage perturbations under the first excitation, while $U_{\text{d}2}$ and $U_{\text{q}2}$ correspond to those under the second excitation. The resulting current perturbations are $I_{\text{d}1}$ and $I_{\text{q}1}$, and $I_{\text{d}2}$ and $I_{\text{q}2}$, respectively. Hence, the elements in the MIMO impedance matrix $Z$ can be identified. However, this approach can be time-consuming and prone to inaccuracies if the operating point of the IBR changes during measurements. To address this issue, multi-frequency signal injections such as pseudo-random binary sequence (PRBS) signals \cite{Gong_DQ} and Gaussian Pulse signal \cite{Fan_DQAdmittance_2023} have been proposed, enabling faster and more reliable impedance measurements. 

By utilizing these measured impedances, a gray-box model of the IBR can be developed by incorporating partial system knowledge. A standard approach involves employing vector fitting (VF) algorithms to estimate the transfer functions for each individual channel within the impedance matrix \cite{Zhou_A}. The transfer functions are firstly formed by considering the different IBR control blocks (e.g. as represented in Fig. \ref{fig:GrFRM_highorder_loops}) in the different channels, whose numerators and denominators contain unknown parameters \cite{Zhou_GrayBox_2021}. Optimization algorithms are then applied to estimate these parameters so that the resulting transfer function best matches the measured impedance across different operating points. These methods typically embed a prior-knowledge of control structure, which contributes to increase the resulting model interpretability. Nevertheless, if the assumed control structure is inaccurate, this approach can result in significant modeling errors. To mitigate this issue, an alternative method first employs VF to generate an impedance model that approximates the measurements, and a neural network is then used to compensate for the residual errors between the VF results and the actual measurements. This cascaded architecture leverages the strengths of both analytical and data-driven approaches. However, the complexity of the model  increases, and its accuracy may still degrade if the initial control structure deviates noticeably from the actual system \cite{Zhang_GrayBox_2024}.

\subsection{Black-box impedance models of IBRs for small-signal analysis}
Although frequency-domain impedance model of IBRs derived from measurements has black-box nature, the IBR dynamics and impedance profiles will vary with different OPs, as demonstrated in Figs. \ref{fig:eig_static_valid} and \ref{fig:Overall1}. To address this limitation, data-driven surrogate models (e.g., neural-network-based \cite{Li_Neural} or support-vector-machine-based \cite{Mohammed_SVM_2024}) have been proposed to efficiently capture the dynamic behavior of IBRs. A data-driven surrogate model uses machine learning algorithms to learn the mapping between inputs and outputs from a collected dataset, effectively creating a computationally efficient black-box approximation of IBR impedance profiles at different OPs. It should be noted that the neural network surrogate is trained using data collected around stable OP, assuming the existence of a static equilibrium point. Hereafter, OP denotes the inverter setpoint with an existing equilibrium.


\newlist{tabitemize}{itemize}{1}
\setlist[tabitemize]{
    label=\textbullet,
    leftmargin=*,
    align=left,
    rightmargin=0pt,
    nosep,
    topsep=0pt,
    partopsep=0pt,
    parsep=0pt
}

\begin{table*}[t!]
\centering
\caption{Summary and comparison for existing data-driven IBR modeling methods.}
\label{tab:datadriven_table}
\renewcommand{\arraystretch}{1.0}
\footnotesize
\begin{tabularx}{\textwidth}{
    >{\centering\arraybackslash}m{1.4cm}|  
    >{\centering\arraybackslash}m{1.6cm}|  
    >{\centering\arraybackslash}m{2.0cm}|  
    >{\centering\arraybackslash}m{2.0cm}|  
    >{\raggedright\arraybackslash}p{3.6cm}|  
    >{\raggedright\arraybackslash}p{3.4cm}   
}
\hhline{|------|}
\textbf{Types} &
\multicolumn{2}{c|}{\textbf{Method}} &
\textbf{Application} &
\textbf{Characteristics} &
\textbf{Research gap}
\\ \cline{1-6}

\multirow{8}{*}{\makecell[c]{Gray-\\box\\model}} &
\multicolumn{2}{>{\centering\arraybackslash}m{4.2cm}|}{Vector-fitting-based impedance model \cite{Zhou_A,Zhou_GrayBox_2021,Zhang_GrayBox_2024}} &
Small-signal stability &
\begin{tabitemize}
    \item Medium prior knowledge
    \item Medium-high physical explainability
    \item Low-medium generalizability
\end{tabitemize} &
\begin{tabitemize}
    \item Lack of uniqueness
    \item Unable to adapt to various control diagrams
    \item High model order for complex control or high bandwidth
\end{tabitemize}
\\ \cline{2-6}

&
\multicolumn{2}{>{\centering\arraybackslash}m{4.2cm}|}{Parameter identification \cite{Buttner_Complex-Phase_2025}, \cite{Abdalmoaty_EPSR_2027}, \cite{Haberle_CDC_2023}} &
Large-signal stability &
\begin{tabitemize}
    \item Medium prior knowledge
    \item Medium-high physical explainability
    \item Medium generalizability
\end{tabitemize} &
\begin{tabitemize}
    \item Limited to specific topologies
    \item Requires accurate initial parameter estimations
\end{tabitemize}
\\ \cline{1-6}

\multirow{22}{*}{\makecell[c]{Black-\\box\\model}} &
\multicolumn{2}{>{\centering\arraybackslash}m{4.2cm}|}{EMT model \cite{EMT_mahseredjian,EMT_mahseradjan_2,Smith_Black}} &
\makecell[c]{Large-/Small-\\signal stability} &
\begin{tabitemize}
    \item Low prior knowledge
    \item Low physical explainability
    \item Low generalizability
\end{tabitemize} &
\begin{tabitemize}
    \item Lack of physical interpretability
    \item Time-consuming for multiple OP simulations
    \item Less insight for controller design or stability analysis
\end{tabitemize}
\\ \cline{2-6}

&
\multicolumn{2}{>{\centering\arraybackslash}m{4.2cm}|}{Impedance measurements \cite{Gong_DQ,Wen_Small,Fan_DQAdmittance_2023}} &
Small-signal stability &
\begin{tabitemize}
    \item Low prior knowledge
    \item Low physical explainability
    \item Low generalizability
\end{tabitemize} &
\begin{tabitemize}
    \item Lack of physical interpretability
    \item Limited to linearized operating point
\end{tabitemize}
\\ \cline{2-6}
&
\multirow{10}{*}{\makecell[c]{OP-driven\\neural\\network}} &
Direct mapping \cite{xiongfei_edge,Li_Neural,Zhang_ANN_2021,yicheng_neural,Shen_KAN_2026,Ali_DropoutKAN_2025,Lyu_DataDriven_2024} &
\multirow{10}{*}{\makecell[c]{Small-signal\\stability}} &
\begin{tabitemize}
    \item Low prior knowledge
    \item Low physical explainability
    \item Low-medium generalizability
\end{tabitemize} &
\begin{tabitemize}
    \item Lack of physical interpretability
    \item Unable to adapt to various control structures
    \item Heavy dependence on training data
    \item Lack of uncertainty awareness
\end{tabitemize}
\\ \cline{3-3} \cline{5-6}

& &
Physics-informed mapping \cite{Zhang_Physics-informed_2023,Li_Few-Shot_2025} &
&
\begin{tabitemize}
    \item Medium prior knowledge
    \item Low-medium physical explainability
    \item Medium generalizability
\end{tabitemize} &
\begin{tabitemize}
    \item Inaccurate under various control diagrams
    \item Heavy dependence on training data
    \item Lack of uncertainty awareness
\end{tabitemize}
\\ \cline{2-6}

&
\multicolumn{2}{>{\centering\arraybackslash}m{4.2cm}|}{Feature-driven neural network \cite{yang_imped_datadriven}} &
Small-signal stability &
\begin{tabitemize}
    \item Low prior knowledge
    \item Low physical explainability
    \item Medium generalizability
\end{tabitemize} &
\begin{tabitemize}
    \item Not directly connected with OPs
    \item Lack of uncertainty awareness
\end{tabitemize}
\\ \cline{2-6}

&
\multicolumn{2}{>{\centering\arraybackslash}m{4.2cm}|}{Nonlinear dynamic model \cite{Yu_DeepLearning_2017,Zheng_LFINODE_2026,Zheng_PINODE_2025,Liu_NeuralLyapunov_2025}} &
Large-signal stability &
\begin{tabitemize}
    \item Low prior knowledge
    \item Low-medium physical explainability
    \item Low generalizability
\end{tabitemize} &
\begin{tabitemize}
    \item Unable to adapt to various control diagrams
    \item Negligence of complex inner control
    \item Heavy dependence on training data
    \item Lack of uncertainty awareness
\end{tabitemize}
\\ \cline{1-6}
\end{tabularx}
\end{table*}

\paragraph{\textbf{OP-driven neural-network-based model}}
To link different OPs to IBR impedances, the neural network can take the OPs (e.g., active power $P$, reactive power $Q$, and PCC voltage $V_{\text{pcc}}=V_g$) together with the inspected frequency $f$ as inputs \cite{Li_Neural}, while the real and imaginary parts of the impedance across different channels are used as outputs, as shown in Fig.\ref{fig:NN1}(a). Trained on dataset covering a wide range of OPs\cite{Zhang_Physics-informed_2023}, the neural network could learn the mapping from OPs to IBR impedance as:
\begin{equation}
\mathbf{Z}_{\text{NN-OP}} =
\begin{bmatrix}
Z_{\text{dd}} (P, Q, V_{\text{pcc}}, f) & Z_{\text{dq}} (P, Q, V_{\text{pcc}}, f)\\
Z_{\text{qd}} (P, Q, V_{\text{pcc}}, f) & Z_{\text{qq}} (P, Q, V_{\text{pcc}}, f)
\end{bmatrix}.
\label{eqn:imped_OP}
\end{equation}

The trained neural network then serves as a computationally efficient surrogate model that can predict the impedance across diverse operating conditions without the need for repeated measurements through perturbation injection \cite{Li_Few-Shot_2025},\cite{Li_Sequence_2025}. When experimental data are limited, the neural network can be pre-trained using high-fidelity simulation data and subsequently fine-tuned with a small amount of experimental data, which can achieve acceptable accuracy while significantly reducing the measurement burden \cite{Li_Neural}. Transfer learning also offers an effective solution, realized by strategically adjusting the network structure to leverage weights from pre-trained models for faster convergence and improved accuracy \cite{Zilio_TransferLearning_2024}.

To ensure modeling accuracy, the structure of the neural-network-based surrogate model requires careful design. Research in \cite{yicheng_neural} has demonstrated that, when the transfer function of each channel in the IBR impedance matrix contains \textit{m} poles and \textit{n} zeros, the number of neurons in the hidden layer have a minimum dimension of 2max(\textit{m}, \textit{n})+2 to guarantee sufficient approximation accuracy. In addition, during neural network training, it should be noted that different variables of the OP (see equation \eqref{eqn:imped_OP}) do not equally influence the resulting impedance profile. Taking a GFL-IBR as an example, existing studies have shown that, under certain control and grid configurations, variations in the PCC voltage have a relatively weak impact on the impedance characteristics compared with variations of the active and reactive power. Consequently, during neural network training, a denser sampling of operating points with respect to active and reactive power can be adopted to improve model accuracy and data efficiency, while fewer variations in PCC voltage are required. 

Another approach to enhance modeling accuracy while simplifying the neural network architecture involves integrating physical information into the model \cite{Li_Few-Shot_2025}. A common strategy is to assign physical meaning to individual layers of the network \cite{Zhang_Physics-informed_2023} and train them independently, as shown in Fig.\ref{fig:NN1}(b). By cascading these pre-trained layers, a comprehensive surrogate model can be eventually formed. However, this method requires partial knowledge of the IBR control system, which inevitably compromises the strictly black-box nature of the surrogate.

When properly trained, these methods can serve as accurate impedance modeling tools for IBRs, enabling real-time dynamic analysis of IBR-dominated grids including impedance-based stability assessment and interaction analysis. If the inspected frequencies are fixed, the neural network structure can be further simplified by using only the operating points as inputs. Recently, advanced neural network architectures, such as the Kolmogorov Arnold Network \cite{Shen_KAN_2026}, \cite{Ali_DropoutKAN_2025}, have also been applied to IBR impedance estimation to improve modeling precision. However, these methods may become ineffective for aggregated IBRs, because the one-to-one mapping between the operating point and the impedance may no longer hold when measurements are taken at the PCC point. The computation burden also increases with the number of IBRs \cite{Lyu_DataDriven_2024}. 

\paragraph{\textbf{Feature-driven neural-network-based model}}
To solve the above limitations, \cite{yang_imped_datadriven} has proposed moving the data-driven impedance modeling from direct OP-impedance mappings toward feature-impedance mappings, as shown in Fig.\ref{fig:NN1}(c). A stacked autoencoder has been proposed to extract the low-dimensional features from measured impedance profiles of an IBR system as shown in Fig.\ref{fig:10a}, avoiding the one-to-one dependence on OP-impedance mapping that constrains the OP-driven neural networks. By compressing high-dimensional impedance profiles into a smaller number of features, the effects of controller structure, system parameters and operating points are implicitly captured, while the accompanied decoder can reconstruct full impedance profile based on any combination of features:
\begin{equation}
\mathbf{Z}_{\text{NN-Feature}} =
\begin{bmatrix}
Z_{\text{dd}} (F_1, ..., F_n) & Z_{\text{dq}} (F_1, ..., F_n)\\
Z_{\text{qd}} (F_1, ..., F_n) & Z_{\text{qq}} (F_1, ..., F_n)
\end{bmatrix}
\end{equation}
where $F_i$ represents the $i$th feature extracted. Therefore, although only using impedance measurements under stable OPs, unstable OPs for the IBR-grid system can also be predicted by enlarging the feature space. Such capability addresses a long-standing limitation for online impedance prediction, where unstable OPs in small-signal perspective are inherently unmeasurable in practice.

By conducting thorough searching over the enlarged feature space, a cluster of probable impedance profiles for an IBR system can be obtained. For a given grid impedance, the stability boundary of the IBR can then be derived by performing stability evaluation on each predicted impedance profile using generalized Nyquist criterion, as shown in Fig. \ref{fig:10b}. This process facilitates efficient stability assessment based on single-frequency impedance measurement. From a system-level perspective, feature-driven neural networks exhibit several key advantages. First, because the impedance is mapped from latent features rather than explicit OPs, the impedance prediction remains valid for multi-IBR systems in which the measured  PCC impedance may vary due to internal power distributions. Second, low-dimensional feature representation can reduce the computational burden of impedance prediction while maintaining high accuracy. For large-scale integration of IBRs from different vendors, the proposed feature-driven framework inherently suppresses redundant and irrelevant information by projecting high-dimensional impedance data onto a compact latent space, thereby enabling scalable and effective impedance prediction without detailed knowledge of individual controller implementations.

\subsection{Black-box nonlinear model of IBRs for large-signal analyses}

To capture large-signal dynamics of IBRs for transient stability analysis, recent research has increasingly employed data-driven techniques to construct nonlinear dynamic models of IBRs considering unavailable control details \cite{Yu_DeepLearning_2017}. \cite{Liu_Dissipation_2026} proposes treating the GFM-IBR as an unknown nonlinear dynamical system with fixed order, and a feedforward neural network is trained to approximate the full nonlinear behavior directly from measurement data. This learned nonlinear model preserves the continuous-time structure of IBR system and enables large-signal transient stability analysis through a neural-approximated Lyapunov function, estimating the region of attraction (see Fig. \ref{fig:stability-ball}(c)) without requiring explicit analytical equations. \cite{Zheng_LFINODE_2026} proposes coupling neural networks trained independently on large-signal and small-signal dynamics to enhance overall modeling efficiency and accuracy.

Overall, these works demonstrate a recent shift from black-box linearized IBR representations toward direct nonlinear system identification, where neural networks approximate the underlying differential equations of IBR dynamics \cite{Zheng_PINODE_2025}. 
However, developing such a data-driven nonlinear model of IBRs is considerably more challenging than a small-signal linearized model. First, large-signal modeling must capture full nonlinear dynamics across wide operating ranges (including current limitation), demanding far larger and more diverse training datasets and imposing a significantly heavier computational burden. Moreover, large-signal behavior is strongly influenced by grid interactions and physical constraints, introducing complex and non-smooth characteristics difficult to capture in training data. Additionally, large-signal transients span multiple time scales simultaneously, making the data-driven models far less generalizable and more expensive to develop. 
Overall, identifying data-driven nonlinear IBR models able to fully capture the interplay among static, small- and large-signal phenomena as discussed in Section \ref{sec:model-based} remains still an open issue.
\subsection{Data-driven stability analysis and data-driven control }
Data-driven stability analysis has emerged as an effective complement to modeling approaches for IBRs, particularly in scenarios where large-scale operating condition screening is required. 
Instead of learning impedance models, neural-networks could directly map OPs to stability labels \cite{Hossen_SafeOperation_2023}, enabling rapid screening of IBR-grid interactions with high computational efficiency. However, a key limitation of conventional neural networks lies in the lack of reliably and the ability to warn prediction failures \cite{HumblotRenaux_UncertaintyAware_2025}. 

A more reliable alternative grounds data-driven stability analysis on control theory, raising the question of which model-based methods in Section \ref{sec:model-based} have data-driven counterparts. For small-signal analysis, linear systems can be represented directly by the input-output trajectories via Willems' Fundamental Lemma, enabling stability analysis directly from data \cite{Markovsky_CSM_2023}. For large-signal stability, physics-informed neural networks (PINNs) offer a promising route by building Lyapunov functions from data \cite{Zheng_LFINODE_2026,karampinis2027neural}. Koopman operator-based methods, which lift nonlinear dynamics into infinite-dimensional linear representations learned from data, can be conceptually leveraged for large-signal analysis too \cite{Choi_Koopman_IBR_2024}.

Another representative line of work focuses on data-driven stability inference based on time-domain measurements \cite{Thor_Probabilistic_2025}, which identifies oscillation modes directly from measured time-domain signals. Both large-signal \cite{Sepehr_PSC_2022} and small-signal \cite{Qin_StabilityMonitoring_2025} instabilities can be potentially detected at an early stage based on measured IBR terminal voltage and currents. Furthermore, by conducting dynamic mode decomposition, fast extraction of modal frequency and damping information can be realized \cite{Kong_DataLight}, enabling real-time stability assessment even under black-box conditions. This class of methods is particularly suited for online monitoring of instability events, as it relies solely on locally measurable signals and avoids time-consuming frequency scanning. \par 
When it comes to data-driven control design, a canonical split exists between indirect (model-based) and direct (model-free) approaches. Whether to pursue a model-based approach or not depends on the specific scenario, and we refer to \cite{dorfler2023data} for a detailed discussion of the merits of both approaches. Whereas models offer interpretability and lend themselves to powerful model-based design techniques, direct model-free design methods can (and often do) outperform indirect model-based techniques. A few reasons are as follows: the best model fitting the data may not be the best model for the intended control design task. Further, it is hard to propagate the uncertainty from the data (variance error) and model selection (bias error) through system identification and to model-based control design. Hence, the model-based design approach is inherently compromised by uncertainty, and that is why the direct model-free approach is increasingly pursued. While there is a flurry of activities in the control and reinforcement learning communities, we mention mature approaches based on behavioral system theory \cite{Markovsky_CSM_2023,leng2026data,zhao2025direct,huang2021decentralized,carlet2022data} that have reached the power systems/electronics implementation stage. Most of these methods rely on predictive control and thus real-time optimization. Hence, their applications are naturally restricted by real-time computability and have only recently become feasible to implement. The theory and computer implementations of data-driven control are so far restricted to linear world models (an example code repository can be found in \cite{datadriven_repository}), nevertheless further advancements are to be expected. \par
A summary for existing data-driven IBR modeling methods is presented in Table \ref{tab:datadriven_table}. Despite the recent promising advances in data-driven modeling for IBRs, several critical challenges remain to be addressed. First, generalizability remains a fundamental concern for data-driven methods. Models trained on data from specific operating conditions and inverter configurations often struggle to perform reliably when adopted under untrained inverter or grid conditions. Second, multi-inverter operation in future IBR-dominated grids introduce more complexity, as interactions between parallel or networked inverters give rise to emergent and unseen dynamics that training data from single IBR cannot adequately capture. Third, data silos existing among different manufacturers and grid operators (see Fig. \ref{fig:GrFRM_highorder_loops}) pose practical barriers to large-scale data sharing, particularly in industrial settings. Open-source and comprehensive dataset for IBR modeling remains lacking at present. Fourth, the limited incorporation of physical knowledge indicates that purely data-driven models may violate fundamental physical laws or constraints in real IBR operation, reducing their trustworthiness in safety-critical applications. The interpretability (or lack thereof) of complex neural network models also fails to provide insights for control decisions.

\section{Conclusions and Future Trends} \label{sec:concl}
This paper has discussed the current challenges and recent advancements in model-based and data-driven stability analysis of power electronics-dominated power grids.
Three kinds of converter-driven instability -- namely static, small-signal, and large-signal dynamic instability -- have fist been defined, and the main factors influencing each of them have been discussed (see Table \ref{tab:physic_constraints}). The manuscript has demonstrated the strict \emph{interdependence} among these three types of instability, and stressed the necessity of studying them concurrently through proper stability analysis tools. 
In the light of this result, model-based tools for static, small-signal and large-signal stability analysis have been examined, and their capability of investigating different instability scenarios has been assessed (see Figs. \ref{fig:interplay1} and \ref{fig:interplay2}).
Afterwards, data-driven tools for stability analysis and control have been explored, motivated by the unattainability of detailed IBRs and power grid models in the presence of IP-protection, and their application potential and research gaps have been debated (see Table \ref{tab:datadriven_table}). Open-source code repositories for model-based and data-driven stability analysis have been mentioned \cite{repository_cecati_nonlinmod,datadriven_repository,xiongfei_edge}.   \par 
To conclude, the authors identify the following research gaps that deserve further investigation:
\begin{itemize}
    \item Analytical IBR models used for large-signal analysis are often oversimplified with respect to their small-signal counterpart, mainly due to the low scalability of transient stability analysis methods. Large-signal analysis in multi-IBR power grids and considering physical constraints (e.g. current limitation, energy source limitations, etc.) require further investigation.
    \item Decentralized criteria to ensure small-signal stability and interoperability among IBRs in multi-vendor power systems are currently under investigation. Nevertheless, there is still insufficient research on the determination of IBR functional specifications to ensure large-signal stability.
    \item While data-driven small-signal analysis is reaching a good level of maturity, data-driven models for large-signal analysis leave key aspects unaddressed. Embedding prior-knowledge in the model structure is considered a promising solution, nevertheless its repercussions on the model generalizability still represent an obstacle.
    \item In power grids where multiple IBRs implement data-driven control strategies in a decentralized fashion, little understanding of their interaction mechanism and limited results on stability guaranties currently exist.
\end{itemize}
\bibliographystyle{Bibliography/IEEEtranTIE}
\bibliography{Bibliography/IEEEabrv,Bibliography/bibliography}\ 

\end{document}